\documentclass[fleqn,10pt]{wlscirep}

\usepackage[english]{babel}
\usepackage[utf8]{inputenc}
\usepackage{amssymb}
\usepackage{amsmath}
\usepackage{color}
\usepackage{hyperref}
\usepackage{bbm}
\usepackage{epsfig}

\hypersetup{
    bookmarks=false,        
    colorlinks=true,        
    linkcolor=blue,        
    citecolor=blue,        
    filecolor=blue,      
    urlcolor=blue
}

\newcommand{\be}{\begin{equation}}
\newcommand{\ee}{\end{equation}}
\newcommand{\bea}{\begin{eqnarray}}
\newcommand{\eea}{\end{eqnarray}}

\newcommand{\fig}[1]{Fig.~\ref{#1}}

\newcommand{\e}{\varepsilon}
\newcommand{\w}{\omega}
\newcommand{\s}{\sigma}
\newcommand{\G}{\Gamma}
\newcommand{\up}{\uparrow}
\newcommand{\down}{\downarrow}

\newcommand{\T}{\mathcal{T}}

\newcommand{\TK}{T_{\rm K}}
\newcommand{\exch}{\Delta\varepsilon_{\rm exch}}

\newcommand{\new}[1]{{ \color{black} #1 }}

\usepackage{ulem}
\usepackage{verbatim} 

\title{Spin Seebeck effect of correlated magnetic molecules}

\author[1,*]{Anand Manaparambil}
\affil[1]{Institute of Spintronics and Quantum Information, Faculty of Physics, Adam Mickiewicz University in Pozna{\'n}, 
	ul.~Uniwersytetu Pozna\'nskiego 2, 61-614 Pozna{\'n}, Poland}
\affil[*]{anaman@amu.edu.pl}

\author[1]{Ireneusz Weymann}
\affil[1]{Institute of Spintronics and Quantum Information, Faculty of Physics, Adam Mickiewicz University in Pozna{\'n}, 
	ul.~Uniwersytetu Pozna\'nskiego 2, 61-614 Pozna{\'n}, Poland}

\date{\today}

\begin{abstract}
In this paper we investigate the spin-resolved thermoelectric 
properties of strongly correlated molecular junctions in the linear response regime.
The magnetic molecule is modeled by a single orbital level
to which the molecular core spin is attached by an exchange interaction.
Using the numerical renormalization group method
we analyze the behavior of the (spin) Seebeck effect,
heat conductance and figure of merit for
different model parameters of the molecule.
We show that the thermopower strongly depends
on the strength and type of the exchange interaction
as well as the molecule's magnetic anisotropy.
When the molecule is coupled to ferromagnetic leads,
the thermoelectric properties reveal an interplay
between the spin-resolved tunneling processes
and intrinsic magnetic properties of the molecule.
Moreover, in the case of finite spin accumulation in the leads,
the system exhibits the spin Seebeck effect. 
We demonstrate that a considerable spin Seebeck effect
can develop when the molecule exhibits an easy-plane magnetic anisotropy,
while the sign of the spin thermopower depends on the type and
magnitude of the molecule's exchange interaction.
\end{abstract}

\begin{document}

\flushbottom
\maketitle

\pagestyle{empty}

\section*{Introduction}

Thermoelectric properties of nanoscale systems have recently attracted
a considerable attention \cite{Giazotto2006Mar,Szczech2011Mar,Heremans2013Jul,Sanchez2014Nov,Benenti2017Jun}.
This is due to the fact that such systems are expected to offer
much better thermoelectric efficiency as compared to their
bulk counterparts \cite{Hicks1993May,Hicks1993Jun,Mahan1996Jul}.
Moreover, it turns out that studying the behavior of thermoelectric coefficients
can provide additional information about various correlations and quantum interference
present in the system\cite{Beenakker1992Oct,Scheibner2005Oct,Reddy2007Mar,
	Dubi2011Mar,Trocha2012Feb,Thierschmann2016Dec,Jaliel2019Sep,Kleeorin2019Dec}.
One prominent example resulting from electronic correlations is the Kondo effect \cite{Kondo_Prog.Theor.Phys32/1964,Hewson_book}
observed in quantum dots and molecules \cite{Goldhaber_Nature391/98,Cronenwett_Science281/1998},
for which the sign changes of the thermopower have been proposed as additional 
signatures and measures of the strength of Kondo correlations \cite{Costi2010Jun,Tooski2014May,Wojcik2016Feb,Nguyen2020Jul}.
In fact, thermoelectric properties of Kondo-correlated nanoscale junctions
have been recently explored experimentally \cite{Scheibner2005Oct,Svilans2018Nov,Dutta2019Jan}. 

Interestingly, the thermoelectric phenomena of nanostructures
have also been explored in the case of systems involving magnetic components.
In fact, with the discovery of the spin Seebeck effect \cite{Uchida2008Oct},
a new field of interest, namely spin caloritronics, have started blossoming \cite{Johnson2010Mar,Bauer2012May,Yu2017Mar,Back2019Mar,uchida_transport_2021}.
It transpires that the interplay of charge, heat and spin  
gives rise to a rich behavior of the thermoelectric coefficients that are now spin-dependent \cite{Krawiec2006Feb,Dubi2009Feb,Swirkowicz2009Nov}.
In correlated magnetic nanostructures, such as e.g. quantum dots coupled to ferromagnetic leads, the spin thermopower
was shown to provide further information about an exchange field
and its interactions with electronic correlations driving the Kondo effect \cite{Weymann2013Aug,Weymann2016Jan}.
Moreover, the spin Seebeck effect has also been
studied in the case of quantum dots subject to external magnetic field \cite{Rejec2012Feb,Costi2019Oct,Costi2019Oct2}.
An interesting situation occurs when the junction comprises 
a molecule of large spin, since the spin Seebeck effect is then additionally
conditioned by intrinsic parameters of the molecule, such as an exchange interaction,
magnetic anisotropy or the magnitude of the molecule's spin \cite{Koch_Phys.Rev.B70/2004,Wang2010Jul}.
In fact, the spin-dependent thermoelectric properties of large-spin molecular junctions
have already been studied in the case of weak coupling to the contacts \cite{Misiorny2014Jun,Misiorny2015Apr,Niu2018Nov,Hammar2019Mar},
whereas the system's behavior in the strongly correlated case remains to a large extent unexplored.
This comprises the goal of this paper, which is to further extend the understanding of thermoelectricity in
strongly correlated magnetic molecular systems.

We therefore undertake the studies of the Seebeck and spin Seebeck effects
for a large-spin molecule, such as a single molecular magnet \cite{Gatteschi_Science265/1994,Gatteschi_book,Gatteschi_Angew.Chem.Int.Ed.42/2003},
embedded in a tunnel junction with either nonmagnetic or ferromagnetic contacts. 
The focus is on the linear response regime with respect
to the applied potential and temperature gradients, which 
justifies the usage of the numerical renormalization group (NRG) method \cite{Wilson_Rev.Mod.Phys.47/1975,Bulla_Rev.Mod.Phys.80/2008}
for the calculations. This method allows for obtaining very accurate results for the electrical and heat conductances
as well as the Seebeck effect and the corresponding figure of merit 
in the full parameter space of the model.
The molecule is assumed to possess an orbital level, through
which transport takes place, which is exchange coupled to 
the spin of the molecule's internal core \cite{Elste_Phys.Rev.B73/2006,Timm_Phys.Rev.B73/2006,Misiorny_Phys.Rev.Lett.106/2011}.
First of all, we show that the Seebeck coefficient
strongly depends on the type and strength of the exchange interaction.
In particular, for antiferromagnetic exchange, we find
an additional sign change of the thermopower as a function of temperature.
Moreover, in the case of magnetic contacts, the interplay
of Kondo screening with exchange field determines the thermoelectric
response of the system. 
We consider two specific cases regarding the spin relaxation time in the leads \cite{Swirkowicz2009Nov,Weymann2013Aug}.
In the case of slow spin relaxation, a spin bias can be generated in the system,
which gives rise to the spin Seebeck effect. On the other hand, for fast spin relaxation,
the spin Seebeck effect does not develop, however, the thermoelectric coefficients
still exhibit interesting spin-resolved properties due to spin dependence of tunneling processes.
We believe that our study, by providing a comprehensive analysis of (spin) thermopower in the case of large-spin molecules,
adds a new insight into the interplay of heat, charge and spin in magnetic molecules,
contributing thus to further development of molecular spin caloritronics.

\section*{Results}
\label{sec:modelandmethod}

\begin{figure}
	\centering
\includegraphics[width=0.7\columnwidth]{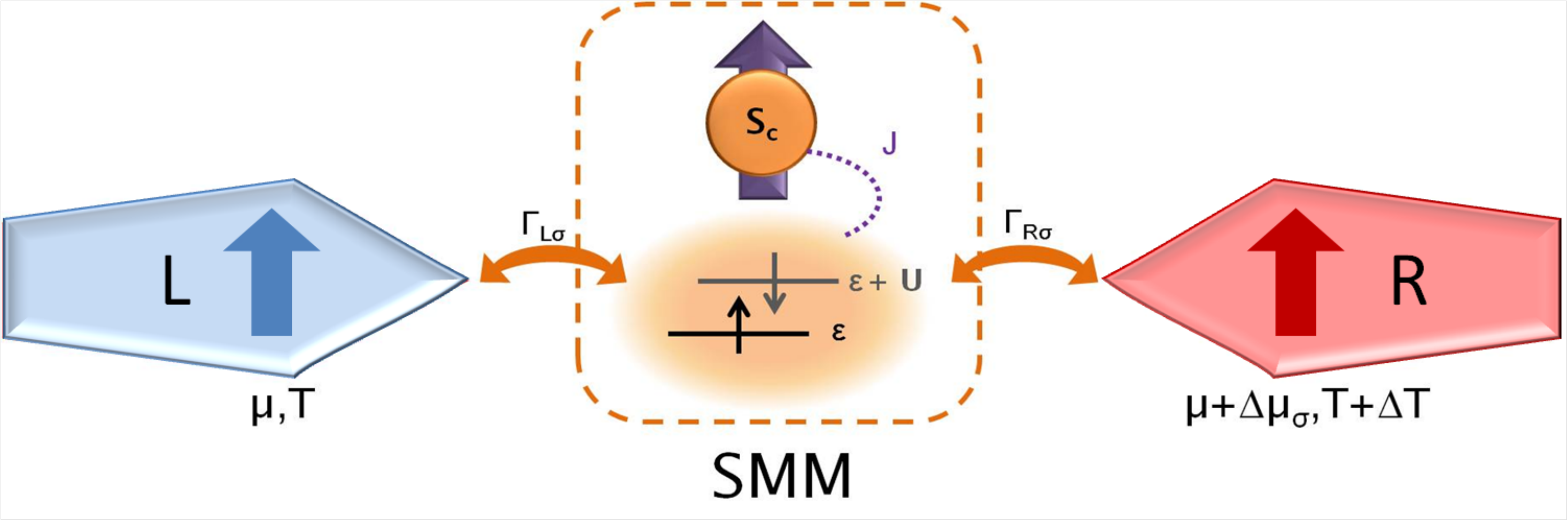}
\caption{
	{\bf  Schematic of the considered molecular junction.}
It consists of a magnetic molecule of spin $S=S_c+s$
tunnel-coupled to external leads,
with spin-dependent coupling strengths, $\Gamma_{L\s}$
and $\Gamma_{R\s}$, for the left and right lead.
$S_c$ is the spin of the molecule's magnetic core,
while $s$ denotes the spin of electrons occupying the orbital level.
The molecule is assumed to effectively
possess one orbital level, through which transport takes place,
that is exchange-coupled (with coupling strength $J$)
to the molecule's core spin $S_c$. 
The orbital level is characterized by on-site energy $\e$ and
the Coulomb correlations $U$.
There is a voltage ($\Delta\mu_\s$) and temperature ($\Delta T$) gradient
applied to the system. In the case of magnetic contacts,
the voltage gradient may be spin-dependent.
\new{In considerations we assume $S_c = 1$,
	while $s=1/2$ ($s=0$) when the orbital level
	occupancy is odd (even).}
}
\label{Fig:schematic}
\end{figure}

The schematic of the studied system is presented in \fig{Fig:schematic}.
A high-spin magnetic molecule, of spin $S=S_c+s$, is coupled to external magnetic leads,
whose magnetizations point in the same direction.
The molecule's easy axis is assumed to coincide
with the direction of leads' magnetizations.
It is further assumed that a single molecular energy level is active in transport
and this orbital level is coupled through an exchange interaction $J$
to the core spin $S_c$ of the molecule.
Thus, the Hamiltonian of the molecule reads as \cite{Elste_Phys.Rev.B73/2006,Timm_Phys.Rev.B73/2006,Misiorny_Phys.Rev.Lett.106/2011},
\be \label{Eq:Hmol}
\hat{H}_{\text{mol}} = \e \sum_\s \hat{n}_\s + U\hat{n}_\up\hat{n}_\down - J\mathbf{\hat{S}_c\cdot\hat{s}} - D \hat{S}_z^2,
\ee
where $\e$ and $U$ denote the energy of the molecule's orbital level and Coulomb correlation energy between two electrons of opposite spin occupying that level.  $\hat{n}_\s\equiv\hat{d}_\s^\dag \hat{d}_\s$ is the occupation number operator for an electron of spin $\s$ and $\hat{d}_\s^\dag$ ($\hat{d}_\s$) is the corresponding creation (annihilation) operator. The spin operator for an electron occupying the orbital level is denoted by $\mathbf{\hat{s}}\equiv(1/2)\sum_{\s\s'}\hat{d}_\s^\dag\mathbf{\s}_{\s\s'}\hat{d}_{\s'}$,
where $\mathbf{\s}\equiv(\s^x,\s^y,\s^z)$ denotes the vector of the Pauli matrices, and 
$\mathbf{\hat{S}_c}$ is the operator for the core spin of the molecule.
The two spins are coupled by the exchange interaction $J$,
which can be either of ferromagnetic ($J>0$) or antiferromagnetic ($J<0$) type,
depending on the sign of $J$.
The molecule can be subject to magnetic anisotropy denoted by $D$
and $\hat{S}_z$ is the $z$th component of the molecule spin operator $\mathbf{\hat{S}}=\mathbf{\hat{S}_c}+\mathbf{\hat{s}}$.
\new{In calculations, we assume $S_c=1$, while the spin of electrons on the orbital level
is given by $s=1/2$ or $s=0$, depending on its occupancy. Consequently,
the total molecule's spin is $S=3/2$ for singly occupied orbital level or  $S=1$ in the case when the occupation is even.
}

The tunneling processes between the molecule and
the leads are described by the following Hamiltonian
\be
  \hat{H}_{\text{tun}} = \sum_{qk\s} v_{qk\sigma} (\hat{c}_{qk\s}^{\dagger} \hat{d}_{\s} + \hat{d}_\s^\dag \hat{c}_{qk\s}),
\ee
where $q={\rm L}$ for the left and $q={\rm R}$ for the right
electrode, the operator $\hat{c}_{qk\s}^{\dagger}$ ($\hat{c}_{qk\s}$) creates (annihilates) an electron with energy $\e_{qk\s}$, momentum $k$ and spin $\s$ in the $q$-th lead, and $v_{qk\sigma}$ denotes the corresponding tunnel matrix elements.
The leads are described within the non-interacting quasi-particle
approximation by
\be
\hat{H}_{\text{leads}}=\sum_{qk\s} \e_{qk\s} \hat{c}_{qk\s}^{\dagger} \hat{c}_{qk\s}.
\ee
Having defined the three parts of the Hamiltonian,
the total Hamiltonian is given by,
$\hat{H} = \hat{H}_{\text{mol}}+ \hat{H}_{\text{tun}}+\hat{H}_{\text{leads}}$.

The tunnel coupling between the molecule and the leads
gives rise to the broadening of the orbital level,
which can be described by, $\Gamma_{q\s} = \pi\rho_{q\s} v_{q\sigma}^2$,
where $\rho_{q\s}$ is the spin-dependent density of states at the Fermi level in the lead $q$ and we assumed momentum-independent tunnel matrix elements
$v_{qk\sigma}\equiv v_{q\sigma}$. We then define the full broadening function,
which for spin $\sigma$ can be written as,
$\Gamma_\sigma = (1+\eta p) \Gamma$,
where $\Gamma = \Gamma_L + \Gamma_R$
and $\Gamma_{q} = \Gamma_{q\uparrow}+ \Gamma_{q\downarrow}$,
whereas $p$ is the effective spin polarization of the left and right ferromagnetic lead,
$p = (p_L + p_R)/2$, and $\eta=1$ ($\eta=-1$) for
spin-up (spin-down) electrons.

\subsection*{Thermopower in the case of nonmagnetic leads}

We focus on the thermoelectric transport properties of the considered molecular junction in the linear response regime. 
The interesting thermoelectric phenomena happening across the magnetic molecule
can be quantified using the transport coefficients,
such as the electrical conductance $G$, the thermopower (Seebeck coefficient) $S$,
as well as the thermal conductance $\kappa$ and the thermoelectric figure of merit $ZT$.
In the linear response regime, these quantities can be expressed
in terms of Onsager integrals $L_{n\s}$ \cite{Onsager_Phys.Rev.38/1931}
\be
L_{n\s} = -\frac{1}{h}\int d\w(\w-\mu)^{n} \frac{\partial f}{\partial \w} \T_{\s}(\w),
\ee
where $\T_\s(\w)$ is the energy-dependent transmission
coefficient for the spin channel $\s$,
$f$ is the Fermi-Dirac distribution function and $\mu$ denotes the electrochemical potential.
The electrical conductance $G$ and the electronic contribution
to the thermal conductance $\kappa$ can be then found from
\be
G=e^2 L_{0}
\;\;\;\;\; {\rm and} \;\;\;\;\; 
\kappa=\frac{1}{T} \left( L_{2} - \frac{L_{1}^2}{L_{0}} \right) ,
\ee
where $e$ is the electron charge, $T$ denotes the temperature
and $L_n=\sum_\sigma L_{n\sigma}$.
On the other hand, the thermopower $S$ is defined as
$S=-[\Delta V/\Delta T]_{J=0}$, on the condition of vanishing of the charge current $J$.
Hence, $S$ can be expressed in the form
\be
S=-\frac{1}{|e| T}\frac{L_{1}}{ L_{0}} .
\ee
Having defined $G$, $S$ and $\kappa$, one can obtain the thermoelectric figure of merit
$ZT\equiv GS^2T/\kappa$.

\begin{figure}[t]
	\includegraphics[width=1\columnwidth]{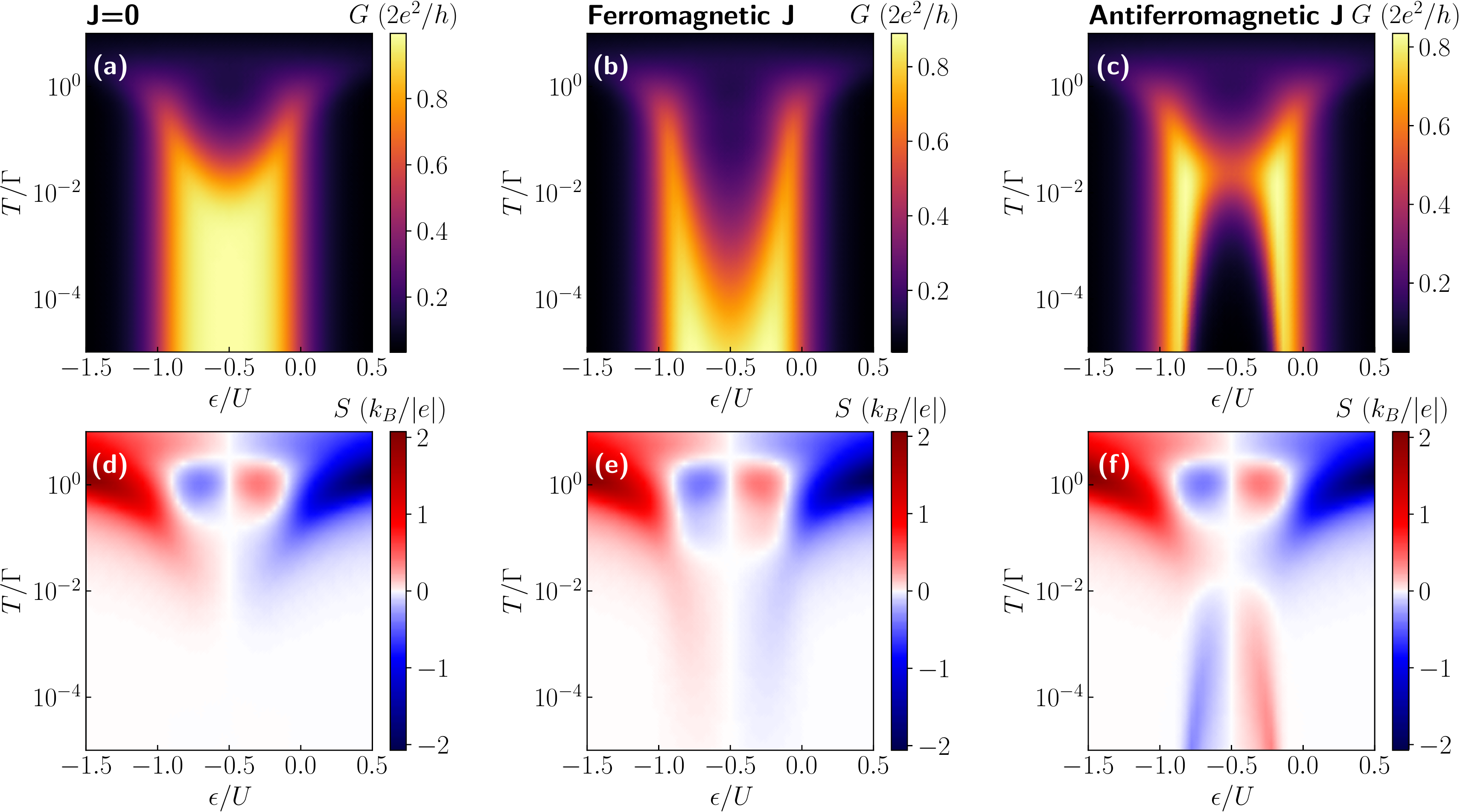}
	\caption{\label{Fig:2DNM}
		{\bf  Conductance and thermopower of molecule in nonmagnetic junction.}
		The linear conductance $G$ and thermopower $S$ as a function
		of the position of molecule's level $\e$ and temperature $T$ 
		in the case of (a,d) $J=0$, (b,e) ferromagnetic ($J=10\TK$) and (c,f) antiferromagnetic ($J=-\TK$)
		exchange interaction, where $\TK = 0.0022$ is the Kondo temperature in the case of $\e=-U/2$, $J=0$ and $p=0$.
		The other parameters are: $\G = 0.05$, $U = 0.5$, $D=0$, in units of band halfwidth, and $p=0$.
		The spin of the molecule is equal to $S=3/2$.
	}
\end{figure}

Let us first consider the case of nonmagnetic molecular junction.
The linear conductance and the Seebeck coefficient as a function
of the molecule's orbital level energy $\varepsilon$ and temperature $T$ are shown in Fig.~\ref{Fig:2DNM}.
The first column shows the results for $J=0$, while the second
(third) column corresponds to the case of the ferromagnetic (antiferromagnetic)
exchange interaction $J$.
The case of $J=0$ is shown just for reference and allows us to
clearly reveal the effects stemming from the presence of large-spin molecule.
To begin with, we consider the behavior of the linear conductance.
The largest changes with lowering the temperature
are visible when the orbital level is singly occupied,
i.e. for $-U \lesssim \e \lesssim 0$. In this case, the Kondo effect can
develop at sufficiently low temperatures, such that $T\lesssim T_K$,
where $T_K$ is the Kondo temperature \cite{Kondo_Prog.Theor.Phys32/1964}. 
Once $T\ll T_K$, the conductance reveals a plateau as a function
of $\e$ of height $G = 2e^2/h$ in the case of $J=0$ \cite{Cronenwett_Science281/1998,Hewson_book}.
However, for magnetic molecules described by the Hamiltonian~(\ref{Eq:Hmol}),
the low-temperature behavior strongly depends
on the type of exchange interaction $J$ \cite{Misiorny2012Jul}. For ferromagnetic 
exchange, the Kondo effect always develops with lowering $T$,
however, $T_K$ becomes reduced compared to the case of $J=0$.
On the other hand, in the antiferromagnetic-$J$ case, once $|J|\gtrsim T_K$,
the spin on the orbital level strongly binds with the magnetic core spin,
which results in the suppression of the conductance through the system.
In this case, in the singly occupied orbital regime, one only
observes a small enhancement (a local maximum) of $G$ with decreasing $T$ followed
by its strong suppression \cite{Misiorny2012Jul}.

The different scenarios discussed above
give rise to a unique behavior of the Seebeck coefficient,
which is presented in the bottom row of Fig.~\ref{Fig:2DNM}.
The first observation is that, as expected \cite{Beenakker1992Oct,Dubi2011Mar}, 
the thermopower changes sign with respect to the particle-hole symmetry point, $\e = -U/2$.
Moreover, one can see that the behavior of $S$ for $T\gtrsim \Gamma$
is hardly affected by the type of exchange interaction $J$.
This is because in our considerations $|J|<\Gamma$
and the effects of finite $J$ can be visible only when $T\lesssim |J|$.
Let us anyway summarize the main features of the high-temperature behavior.
One observes two pronounced maxima for $T\approx \Gamma$
in the case when the orbital level is either empty or doubly occupied. 
On the other hand, when moving to the Coulomb blockade regime where the orbital level
is singly occupied, the thermopower changes sign
and two local extrema develop antisymmetrically around $\e=-U/2$
for $T\approx \Gamma$ \cite{Costi2010Jun,Weymann2013Aug}. With lowering the temperature,
distinct features appear, resulting from the interplay of the correlations driving the Kondo effect
and the molecule's exchange interaction.
First of all, one can note that in the case of $J=0$, the regions
of large $|S|$ extend from the empty and doubly occupied regimes for $T\approx \Gamma$
downwards to low temperatures in the single occupancy regime.
For singly occupied orbital level, the thermopower exhibits
then a sign change as a function of $T$, see \fig{Fig:2DNM}(d),
which signals the relevance of the Kondo correlations \cite{Costi2010Jun}.
A qualitatively similar behavior can be observed in the case
of ferromagnetic exchange interaction $J$, see \fig{Fig:2DNM}(e),
with the main difference associated with smaller temperatures
at which the corresponding sign change occurs.
\new{This is associated with the fact that  the sign change occurs at the onset of the Kondo correlations
and, because the Kondo effect develops at much lower temperatures
in the case of ferromagnetic $J$ compared to the case of $J=0$ [see Figs.~\ref{Fig:2DNM}(a) and (b)],
one observes that the sign change in $S$ is also shifted to lower temperatures. However,
this shift is not proportional to the corresponding shift visible in the behavior of $G$.
It is because while much smaller temperatures
are needed for the full development of the Kondo effect in the case of ferromagnetic $J$,
the temperature associated with the onset of the Kondo correlations
only weakly decreases with $J$. 
This is why the crossover for $J>0$
is only slightly shifted to lower temperatures compared to the case of $J=0$,
cf. Figs.~\ref{Fig:2DNM}(d) and (e).
}
Interestingly, qualitatively new features compared to the case of $J\geq0$ can be observed
in the case of antiferromagnetic exchange interaction,
where an additional sign change at low temperatures is present,
see \fig{Fig:2DNM}(f).

\begin{figure}[t]
	\centering
	\includegraphics[width=0.8\columnwidth]{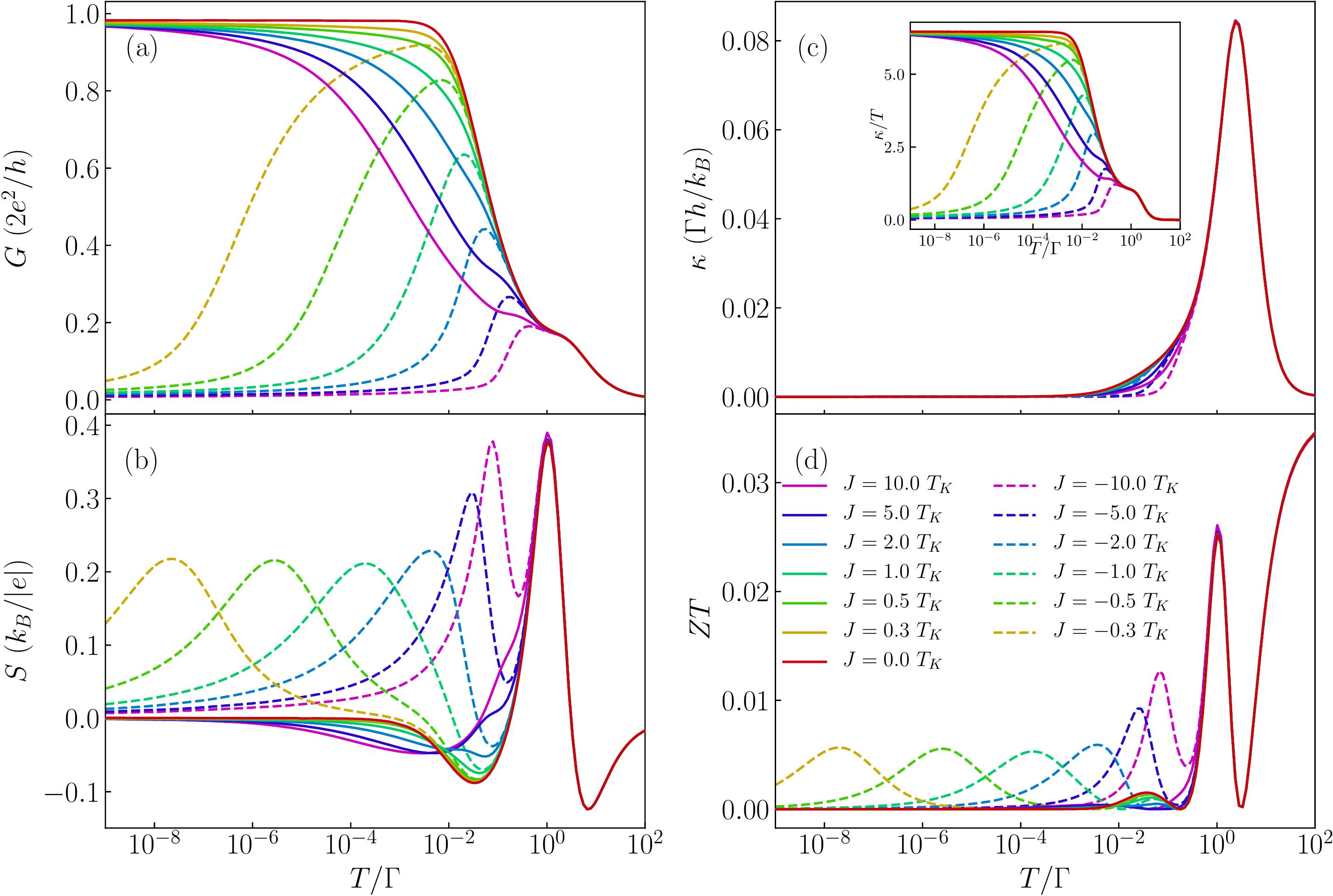}
	\caption{\label{Fig:1DJNM}
		{\bf  Dependence of thermoelectric coefficients on exchange interaction.}
		(a) The conductance, (b) Seebeck coefficient, (c) heat conductance and (d) figure of merit
		as a function of temperature for selected values of exchange interaction $J$
		in the case of nonmagnetic contacts.
		The solid (dashed) lines correspond to the case of 
		ferromagnetic (antiferromagnetic) exchange interaction.
		The inset in (c) presents the temperature dependence of $\kappa/T$.
		The parameters are the same as in \fig{Fig:2DNM} with $\e=-U/3$.
	}
\end{figure}

Further insight into the behavior of the thermoelectric properties can be obtained
from the inspection of \fig{Fig:1DJNM}, which presents the temperature dependence
of $G$, $S$, $\kappa$ and $ZT$ for different values of the exchange interaction $J$, as indicated.
\new{
This figure is generated for the case when the orbital level
is detuned from the particle-hole symmetry point, such that a considerable
thermopower can be observed.
In addition, the orbital level is assumed to be singly occupied ($\e=-U/3$),
such the system is in the local moment regime and the Kondo correlations are relevant at sufficiently low temperatures.}
Let us start with the analysis of the linear conductance.
In the case of ferromagnetic $J$ (see the solid lines in the figure),
the enhancement of $J$ results in a decrease of the Kondo temperature. 
In this case the ground state is always two-fold degenerate with total spin given
by $S = S_c+1/2$ and the Kondo effect develops irrespective of $J$,
though $T_K$ becomes very cryogenic with increasing $J $\cite{Misiorny2012Jul}.
Consequently, we observe mainly quantitative changes in \fig{Fig:1DJNM}(a), while qualitative behavior is the same.
On the other hand, the situation is completely different in the case of antiferromagnetic
exchange interaction $J<0$, see the dashed lines in \fig{Fig:1DJNM}.
Now, with increasing $|J|$, the ground state becomes $S = S_c-1/2$,
since the spin on the orbital level binds anti-ferromagnetically with
the molecule's core spin. Because of that, the Kondo effect is quenched
once the temperature becomes lower than the energy scale responsible
for this antiferromagnetic state. As a result, the temperature dependence
of conductance exhibits a nonmonotonic behavior \cite{Misiorny2012Jul},
see \fig{Fig:1DJNM}(a). We note that such behavior  is similar to the two-stage Kondo effect
observed in side-attached double quantum dots where the hopping induces
an antiferromagnetic interaction between the dots \cite{Wojcik_PhysRevB.91.134422/2014}.

The behavior of the conductance has a strong influence on the Seebeck coefficient.
This is because the temperature dependence of  conductance reflects
the energy dependence of the transmission coefficient $\T(\w)$ and, 
from the Sommerfeld expansion, the thermopower at low temperatures
can be estimated from (note that temperature $T$ is in units of energy)
\be
S \approx - \frac{\pi^2}{3} \frac{k_B}{e} \frac{T}{\T(\w)} \left.\frac{\partial \T(\omega)}{\partial \w} \right|_{\w=0},
\ee
where $\T(\w) = \sum_\sigma \T_{\s}(\w)$.
One can thus see that the thermopower is related to the monotonicity of the variation of the
spectral function with energy.
This is why in the case of ferromagnetic exchange  interaction
we only observe qualitative changes in $S$, see the solid lines in \fig{Fig:1DJNM}(b).
The enhancement of $J$ suppresses then the Kondo temperature,
which is seen in the behavior of $S$ as a shift of the local minimum 
towards smaller temperatures. However,
a completely different scenario develops for antiferromagnetic $J$,
where an additional sign change occurs and $S$ exhibits 
an extra maximum for temperatures corresponding
to the energy scale at which $G(T)$ starts decreasing with lowering $T$,
see the dashed curves in \fig{Fig:1DJNM}(b).

Despite a spectacular impact of exchange interaction on the thermopower,
its effect on the thermal conductance is less pronounced, see \fig{Fig:1DJNM}(c).
This is because $\kappa$ is considerable only at energy scales corresponding to the coupling strength
and Coulomb correlations and thus, as long as $|J|\ll \Gamma$, $\kappa$ hardly
depends on the type and magnitude of exchange interaction.
On the other hand, the figure of merit $ZT$ displays 
new peaks in the case of antiferromagnetic $J$, see \fig{Fig:1DJNM}(d),
which are associated with the above-discussed maxima emerging  in
the temperature dependence of $S$.
We also note that the influence of $J$ on $\kappa$ is more visible
when one plots  $\kappa/T$, see the inset in \fig{Fig:1DJNM}(c).
It is nicely visible that the qualitative behavior of $\kappa/T$ resembles that of the linear conductance,
which is a direct consequence of the Wiedemann-Franz law.
A similar behavior has been recently observed for T-shaped double quantum dots \cite{Wojcik2016Feb}.

\subsubsection*{Effect of magnetic anisotropy}

\begin{figure}[ht]
	\centering
	\includegraphics[width=0.8\columnwidth]{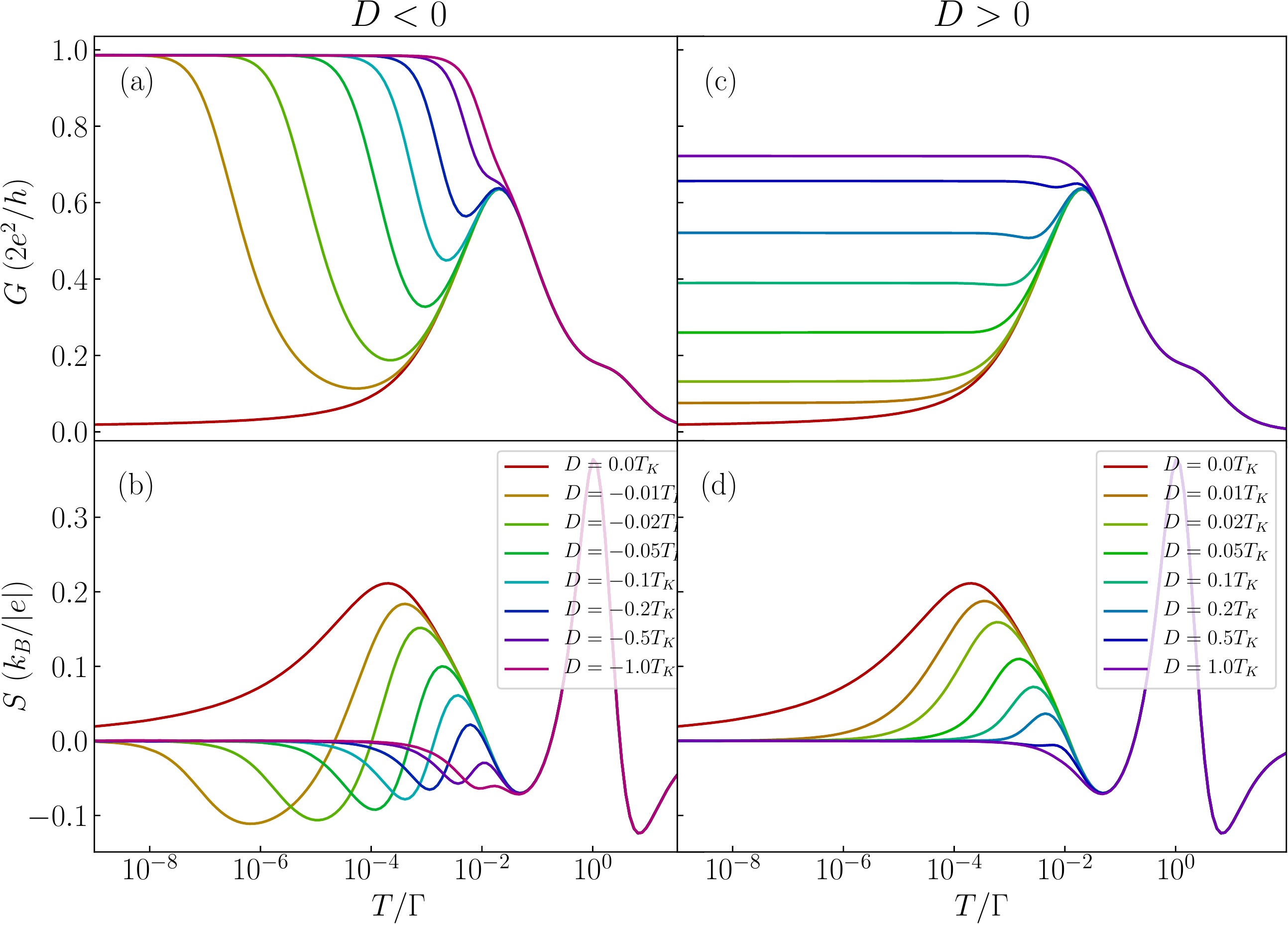}
	\caption{\label{Fig:1DDNM}
		{\bf  The effect of magnetic anisotropy in the case of molecule with nonmagnetic contacts.}
		The temperature dependence of (a,c) the conductance and (b,d) the Seebeck coefficient
		in the case of antiferromagnetic exchange interaction $J=-T_K$
		for selected values of magnetic anisotropy $D$.
		The left (right) column presents the case of easy-plane (easy-axis) type of magnetic anisotropy.
		The parameters are the same as in \fig{Fig:2DNM} with $\e=-U/3$.
	}
\end{figure}

We now focus on elucidating the role of magnetic anisotropy
on the thermoelectric properties of the considered molecular junction.
We consider both an easy-axis ($D>0$) and easy-plane ($D<0$) types of magnetic anisotropy.
First we note that in the case of ferromagnetic exchange interaction between the orbital level
and molecule's core spin the magnetic anisotropy has a very moderate influence
on the thermoelectric properties. It does not lead to new qualitative behavior
as long as $|D|$ is smaller than the corresponding Kondo temperature, therefore 
in the following we just analyze the case of antiferromagnetic exchange coupling $J$.
The temperature dependence of the conductance and the Seebeck coefficient
for this situation is shown in \fig{Fig:1DDNM}, where the left (right) column 
corresponds to the easy-plane (easy-axis) magnetic anisotropy case.
\new{This figure was generated for the same orbital level position as in \fig{Fig:1DJNM},
	such the orbital level is detuned from the particle-hole symmetry point, while
	the system stays in the local moment regime.
}
In the absence of anisotropy the conductance displays a nonmonotonic
dependence, characteristic of the antiferromagnetic exchange coupling.
When an easy plane anisotropy arises in the system and the molecule possesses
a half-integer spin, it results in a two-fold degenerate ground state of $S=1/2$,
such that the Kondo effect can be restored.
This is clearly seen in \fig{Fig:1DDNM}(a), where one observes
an upturn of the conductance with lowering $T$. This in turn has a considerable impact
on the Seebeck coefficient, which exhibits an additional sign change, see \fig{Fig:1DDNM}(b).
On the other hand, in the case of easy plane anisotropy, 
the low-temperature conductance again becomes increased with $D$, 
as can be seen in \fig{Fig:1DDNM}(c).
This is however just associated with a decreased exchange interaction
between the molecule's core spin and the spin of the orbital level,
and not with the reinstatement of the Kondo effect.
Consequently, while the Seebeck coefficient strongly
depends on $D$, no additional sign changes are present.
In fact, the maximum in $S$ for $D=0$ becomes suppressed with increasing $D$
and smears out completely once $D\approx T_K$,
see \fig{Fig:1DDNM}(d).

\subsection*{Thermopower in the case of ferromagnetic leads}

\begin{figure}[t!]
	\includegraphics[width=1\columnwidth]{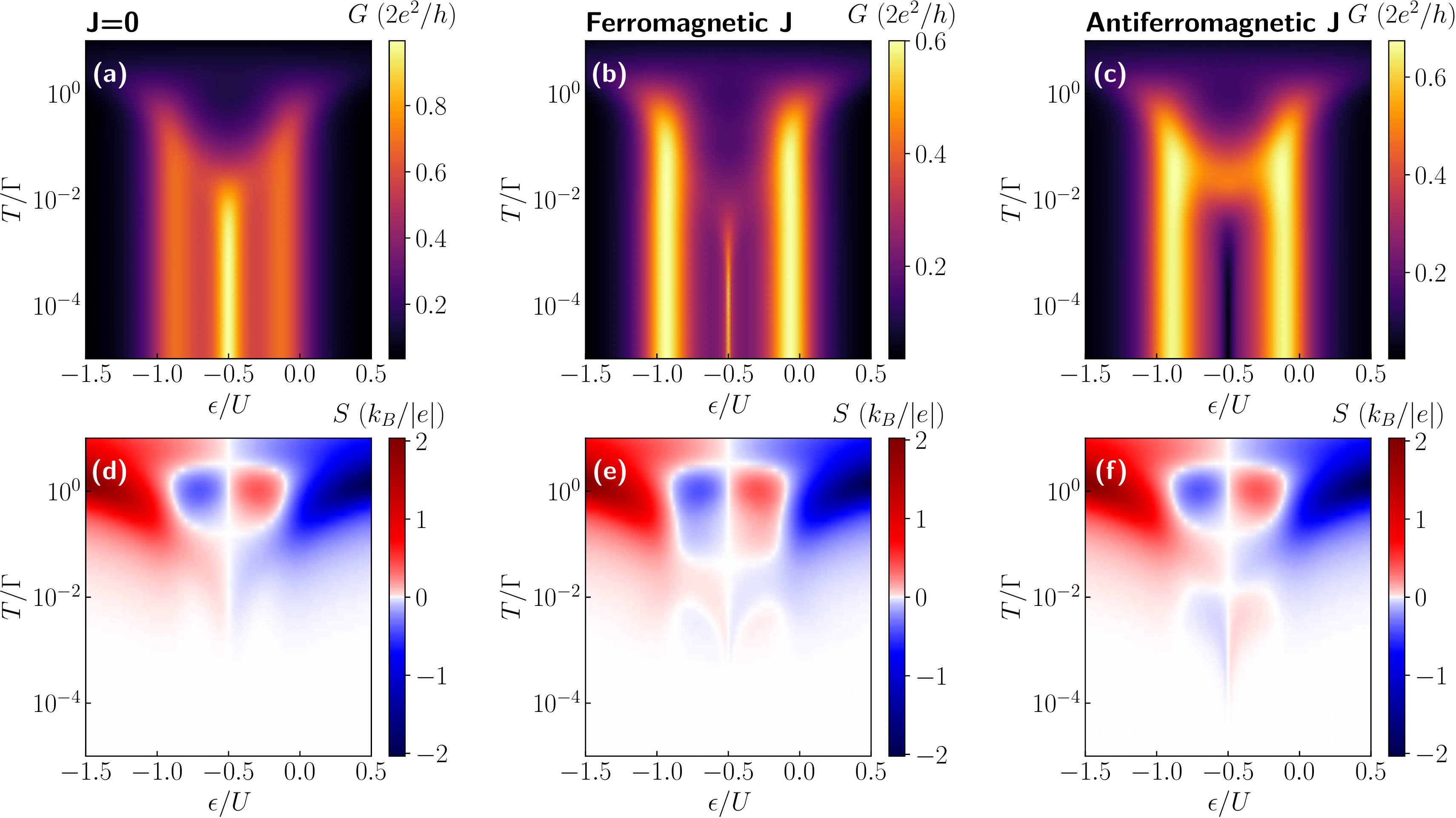}
	\caption{\label{Fig:2DFM}
		{\bf  Conductance and thermopower of molecule in ferromagnetic junction.}
		The linear conductance $G$ and thermopower $S$ as a function
		of the position of molecule's level $\e$ and temperature $T$ 
		in the case of (a,d) $J=0$, (b,e) ferromagnetic ($J=10\TK$) and (c,f) antiferromagnetic ($J=-\TK$)
		exchange interaction. The other parameters are the same as in Fig. \ref{Fig:2DNM} with $p=20\%$.
	}
\end{figure}

We now turn to the discussion of thermoelectric properties
in the case of ferromagnetic electrodes.
For the ferromagnetic contacts, we assume a moderate
spin polarization \cite{Sahoo2005Nov,Merchant_Phys.Rev.Lett.100/2008,Gaass_Phys.Rev.Lett.107/2011}, $p=20\%$.
In this  section we assume that the spin relaxation in the contacts is relatively fast,
such that no spin accumulation develops and the induced potential gradient
does not depend on spin, $\Delta\mu_\uparrow = \Delta\mu_\downarrow$.
In this regime, although the spin Seebeck effect does not develop,
the spin-dependence of tunneling  processes greatly modifies
the thermoelectric transport properties of the system as compared to the nonmagnetic case.
We also note that in the absence of spin accumulation
the formulas for spin-dependent thermoelectric coefficients are the same
as in case of nonmagnetic leads \cite{Swirkowicz2009Nov,Weymann2013Aug}.

The linear conductance and the Seebeck coefficient
as a function of $\e$ and $T$ calculated for different values of $J$
are displayed in \fig{Fig:2DFM}. In the behavior of the conductance
one can clearly observe the signatures of an effective exchange field that develops in the molecule coupled to ferromagnetic electrodes
\cite{Martinek_Phys.Rev.Lett.91/2003_127203,Misiorny_Phys.Rev.Lett.106/2011}.
\new{Such an exchange field, which within the perturbation theory at zero temperature and for $J=0$ can be described as
	\cite{Martinek_Phys.Rev.Lett.91/2003_127203},
	\be
	\exch \approx \frac{2p\G}{\pi}\text{log}\bigg|\frac{\e}{\e+U}\bigg|,
	\ee
	results in a spin-splitting of the molecule's orbital level,
when it is detuned from the particle-hole symmetry point of the model $\e=-U/2$.
In the case of considered molecule, this field depends
in a nontrivial way on the properties of the molecule, such as $J$, $D$ and $S_c$,
however, it still vanishes whenever $\e=-U/2$ \cite{Misiorny_Phys.Rev.Lett.106/2011,Misiorny_Phys.Rev.B84/2011}.}
If the exchange field splitting becomes larger than the Kondo temperature,
it suppresses the Kondo resonance. As a consequence, 
the low-temperature conductance in the Coulomb blockade regime
is generally decreased except for $\e=-U/2$,
where a local maximum as a function of $\e$ is present, see \fig{Fig:2DFM}(a).
A similar conductance suppression can be seen 
in the case of ferromagnetic exchange interaction $J$ presented in \fig{Fig:2DFM}(b).
However, now the suppression of $G$ is larger as compared to the case of $J=0$,
which is associated with a smaller Kondo temperature when $J>0$.
On the other hand, when $J$ is antiferromagnetic,
mainly quantitative changes can be observed in the conductance behavior,
cf. \fig{Fig:2DNM}(c) and \fig{Fig:2DFM}(c).
Namely, the region of suppression of low-temperature conductance is now smaller,
which is an indication that the exchange field hinders the
formation of an antiparallel spin state between the orbital level
and the core spin. Because of that, the decrease
of conductance due to that formation is correspondingly weakened.
The signatures of the interplay of exchange field, Kondo correlations,
and the molecule's exchange interaction are also visible in the behavior
of the thermopower, which is presented in the bottom row of \fig{Fig:2DFM}.
One can see that the main changes are visible in the low-temperature behavior,
which reveals the energy scale associated with the exchange field.
More specifically, in the case of ferromagnetic exchange interaction,
an additional sign change in $S$ occurs as compared to the case of nonmagnetic leads,
whereas for antiferromagnetic $J$ the Seebeck  coefficient
becomes generally reduced, cf. \fig{Fig:2DNM} and \fig{Fig:2DFM}.

\begin{figure}[t]
	\centering
	\includegraphics[width=0.8\columnwidth]{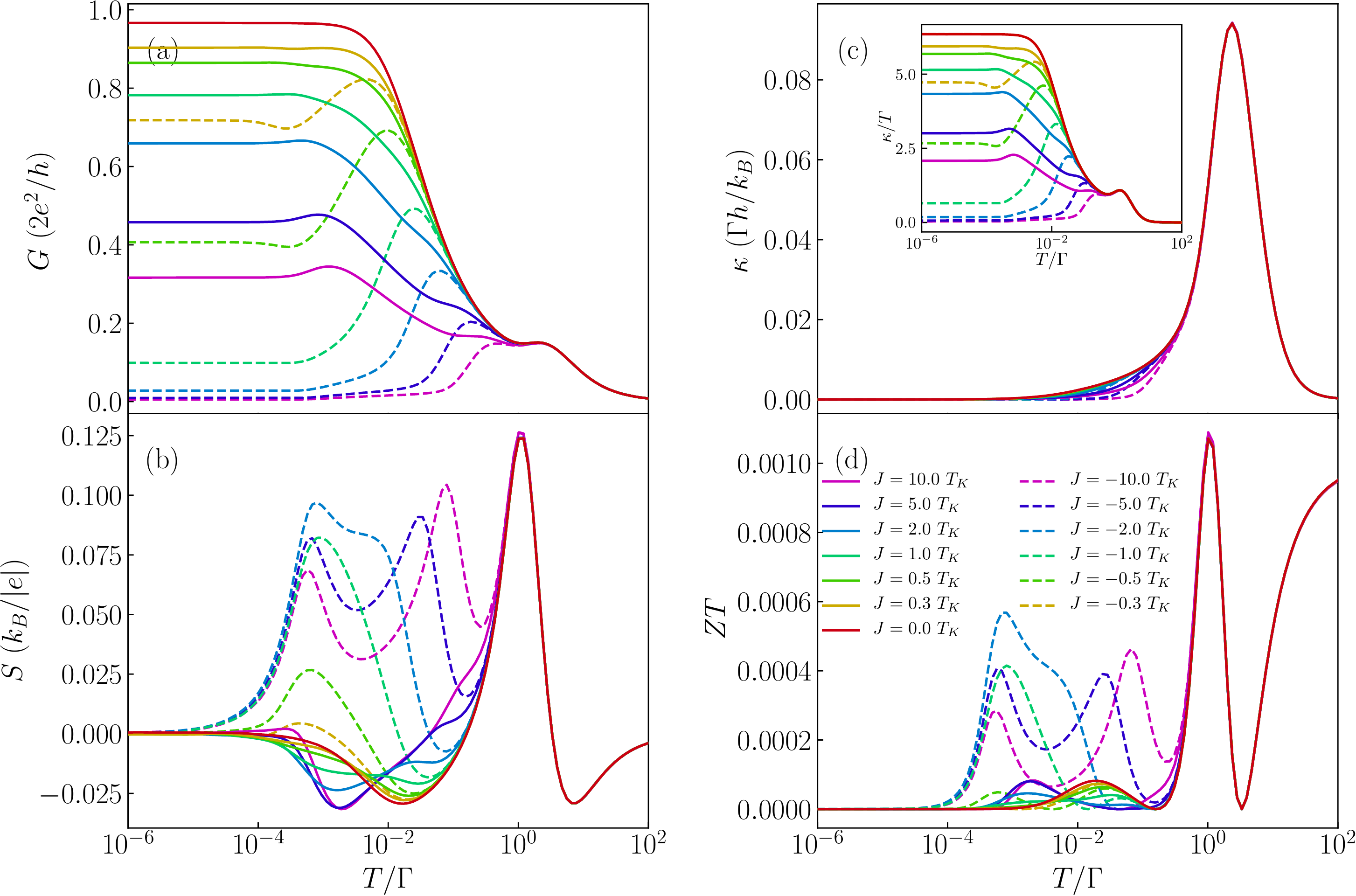}
	\caption{\label{Fig:1DJFM}
		{\bf  Dependence on exchange interaction in the case of ferromagnetic junction.}
		(a) The conductance, (b) Seebeck coefficient, (c) heat conductance and (d) figure of merit
		as a function of temperature for selected values of exchange interaction $J$
		in the case of ferromagnetic leads.
		The solid (dashed) lines correspond to the case of 
		ferromagnetic (antiferromagnetic) exchange interaction.
		The inset in (c) presents $\kappa/T$ as a function of temperature.
		The parameters are the same as in \fig{Fig:2DFM} with $\e=-0.48U$.
	}
\end{figure}
\begin{figure}[t]
	\centering
	\includegraphics[width=0.8\columnwidth]{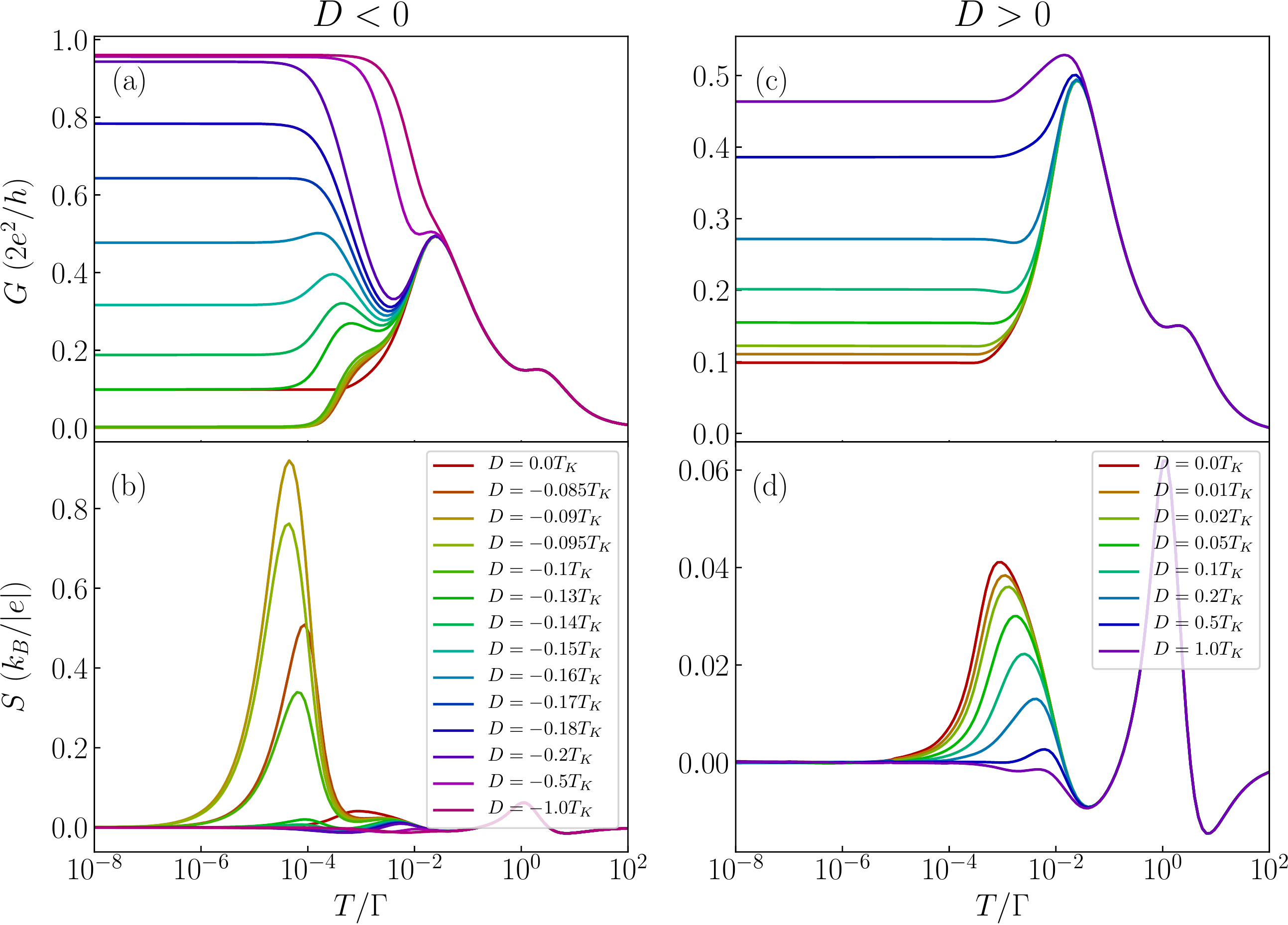}
	\caption{\label{Fig:1DDFM}
		{\bf  The effects of magnetic anisotropy in the case of ferromagnetic junction.}
		The temperature dependence of (a,c) the conductance and (b,d) the Seebeck coefficient
		in the case of antiferromagnetic exchange interaction $J=-T_K$
		for selected values of magnetic anisotropy $D$.
		The left (right) column presents the case of easy-plane (easy-axis) type of magnetic anisotropy.
		The parameters are the same as in \fig{Fig:2DFM} with $\e=-0.48U$.
	}
\end{figure}

To shed more light onto the thermoelectric behavior
of the considered magnetic molecular junction, in \fig{Fig:1DJFM}
we present the temperature dependence of the conductance,
Seebeck coefficient, heat conductance and figure of merit
calculated for different values of exchange interaction, as indicated.
This figure was determined for a relatively low value of detuning
from the particle-hole symmetry point, $\e = -0.48 U$,
such that the exchange field effects rather compete than
surpass other energy scales, and the system's behavior is most interesting.
First of all, one can note that below a certain temperature
the conductance stops changing any more and retains
its low-temperature value. This characteristic energy
scale is set by the exchange field $\exch$---when this field is larger
than thermal energy, it determines the transport behavior
and no further dependence on lowering $T$ is visible, see \fig{Fig:1DJFM}.
The Seebeck effect becomes then suppressed and so does the figure of merit.
This can be understood by referring to the Sommerfeld expansion:
once the exchange field is the dominant energy scale
the conductance becomes constant and so does the low-energy transmission coefficient.
Consequently, $[\partial \T(\omega) / \partial \w ]_{\w=0} \approx 0$ and thus $S\approx 0$.
This explains why $S\approx 0$ and $ZT\approx 0 $ for $T/\Gamma\lesssim \exch/\Gamma \approx 10^{-4}$,
see \fig{Fig:1DJFM}(b) and (d).

Let us now focus on the most interesting behavior, which is present
for temperatures of the order and larger than the exchange field,
and let us start with the case of ferromagnetic exchange interaction $J>0$.
One can see that increasing $J$ results in suppression
of the low-temperature conductance, see \fig{Fig:1DJFM}(a).
This results from the fact that increasing $J$ leads to 
lowering of $T_K$, and once $\exch\gtrsim T_K$, the Kondo peak
becomes suppressed. This is also visible in the thermopower 
and the figure of merit where a new maximum emerges
at energy scale corresponding to the exchange field $T \approx \exch$,
see \fig{Fig:1DJFM}. While in the case of ferromagnetic exchange
interaction mainly qualitative effects are visible, the case
of antiferromagnetic exchange is completely different,
see the dashed lines in \fig{Fig:1DJFM}. In this case,
increasing $|J|$ gives rise to the suppression of the 
conductance due to the formation of an antiferromagnetic spin state
between the molecule's orbital level and its magnetic core.
When $|J|$  is relatively low, the exchange field wins
over the antiferromagnetic interaction and only a small suppression
of $G$ as a function of $T$ is present, see e.g. the curve for $J=-0.3 T_K$ in \fig{Fig:1DJFM}(a).
However, further enhancement of $|J|$, stabilizes
the antiferromagnetic state of the molecule and 
a full suppression of $G$ is obtained once $J\lesssim -2T_K$.
This behavior gives rise to a new maximum visible both in $S$ and $ZT$.
Moreover, while the maximum associated with the conductance
drop as $T$ is lowered moves to higher energies with decreasing $J$ ($J<0$),
there  is an extra maximum visible just at the energy scale
corresponding to $T \approx \exch$.
As a consequence, the temperature dependence of the Seebeck
coefficient displays an interesting triple-peak structure, see \fig{Fig:1DJFM}(b).

To complete the picture, in \fig{Fig:1DDFM} we present
the temperature dependence of $G$ and $S$ calculated 
for different values of magnetic anisotropy. 
Similarly to the case of nonmagnetic leads, we display
the data for antiferromagnetic exchange interaction,
which shows the most interesting behavior. 
Let us first analyze the case of uniaxial anisotropy,
which is presented in the right column of \fig{Fig:1DDFM}.
One can see that finite anisotropy gives rise
to an enhancement of the low-temperature conductance.
This is a consequence of the fact that 
anisotropy breaks the symmetry of the antiferromagnetic state of the molecule
responsible for the conductance suppression.
This effect gives rise to a local maximum in the Seebeck coefficient
that develops at the energy scale
of the order of magnetic anisotropy, see \fig{Fig:1DDFM}(d).
The case when the molecule exhibits easy-plane
type of anisotropy is shown in the left column of \fig{Fig:1DDFM}.
Now, one can observe a very strong dependence 
of both $G$ and $S$ on the magnitude of magnetic anisotropy.
First of all, the low-$T$ conductance exhibits a nonmonotonic dependence on $D<0$.
Once the easy-plane anisotropy is present in the system and $|D|\lesssim 0.1 T_K$, 
$G$ becomes suppressed. However, this tendency becomes reversed
when $D\lesssim -0.1 T_K$, such that one observes an enhancement of 
$G$ at low temperatures, see \fig{Fig:1DDFM}(a).
This is associated with the formation of a doublet ground state
in the molecule in the case of considerable easy-plane anisotropy.
Now, however, one witnesses a subtle interplay
between the antiferromagnetic exchange interaction,
the easy-plane magnetic anisotropy, 
the exchange field that splits the doublet state of the molecule
and the Kondo correlations.
The antiferromagnetic $J$ gives rise to the suppression of $G$
at low temperatures, which is however slightly hindered by the exchange field.
On the other hand, turning on $D$ ($D<0$), results initially in a larger suppression 
of the conductance, nevertheless, increased values of $|D|$
eventually make the doublet state the ground state of the molecule,
enhancing thus $G$ due to the Kondo effect.
Consequently, for sufficiently large $|D|$, the conductance shows
a pronounced Kondo resonance, see \fig{Fig:1DDFM}(a).
The behavior of the conductance is clearly revealed in the 
temperature dependence of the Seebeck coefficient,
which is shown in \fig{Fig:1DDFM}(b).
One can see that for values of $D$ such that the conductance
starts increasing, the thermopower exhibits a considerable
maximum, which actually develops for $T\approx \exch$.
With further increase of $|D|$, this maximum becomes however
decreased and its position moves towards higher temperatures
of the order of the Kondo temperature, see \fig{Fig:1DDFM}(b).
\new{This large enhancement of the Seebeck coefficient
is a result of interplay between the intrinsic properties of the molecule,
such as its magnetic anisotropy, and the ferromagnetism of the leads. Unfortunately,
this effect is not associated with a particularly large figure of merit
since the corresponding electrical conductance is then relatively low,
see \fig{Fig:1DDFM}(a).}

\subsection*{Spin Seebeck effect}

When the electrodes are ferromagnetic and are characterized by a long spin relaxation time,
in addition to the charge current, a spin current can be generated in the system \cite{Swirkowicz2009Nov,Weymann2013Aug}.
The spin current $I_S$ flows if there is a difference between chemical potentials for given
spin direction, i.e. in the presence of a spin bias $\Delta \mu_\uparrow \neq \Delta \mu_\downarrow$.
The development of spin bias is conditioned by the spin relaxation time
in the contacts compared to the time of tunneling events.
The case of fast spin relaxation was discussed in previous section,
now, let us focus on the situation when the spin accumulation can build up in the leads,
i.e. when the spin relaxation time is long.
In such case, the spin Seebeck effect $S_S$ can develop in the system.
Assuming open circuit conditions, i.e. the vanishing of the spin and charge currents,
it can be found from,
\new{
	$S_S = (S_\uparrow - S_\downarrow)/2$, where
	$S_\sigma$ is the thermopower in the spin channel $\sigma$,
	which yields\cite{Swirkowicz2009Nov,Weymann2013Aug}
}
\be
S_S = -\frac{1}{2|e| T}\left( \frac{L_{1\uparrow}}{L_{0\uparrow}}-\frac{L_{1\downarrow}}{L_{0\downarrow}}\right).
\ee
On the other hand, the Seebeck coefficient in the case of finite spin bias is given by
\cite{Swirkowicz2009Nov,Weymann2013Aug},
$S = -\frac{1}{2|e| T}\left( L_{1\uparrow}/L_{0\uparrow} + L_{1\downarrow}/L_{0\downarrow}\right)$,
whereas the heat conductance can be expressed as, 
$\kappa=\frac{1}{T} \sum_\s \left( L_{2\s} - L_{1\s}^2/L_{0\s} \right)$.

\new{
We would like to emphasize that in the case of ferromagnetic leads
in the absence of spin accumulation, the Seebeck effect is due to
a generated voltage difference that is not spin dependent.
However, the tunneling processes themselves do depend on spin,
therefore one then observes the spin-dependent Seebeck effect.
On the other hand, when the spin accumulation is relevant
in the leads, a spin bias becomes generated, which 
gives rise to the spin Seebeck effect. Such spin thermopower results from the spin splitting
of the chemical potentials in the contacts and is associated with the corresponding spin current.
Moreover, we would like to note that spin Seebeck effect due to electron
transport is also referred to as the spin-dependent Seebeck effect,
whereas the typical Seebeck effect in the case of magnetic contacts is referred to as
spin-resolved Seebeck effect \cite{Bauer2012May,uchida_transport_2021}.
However, in our paper we adopt the notion
introduced by \'Swirkowicz et al. \cite{Swirkowicz2009Nov}.
}

\begin{figure}[t]
	\centering
    \includegraphics[width=0.9\columnwidth]{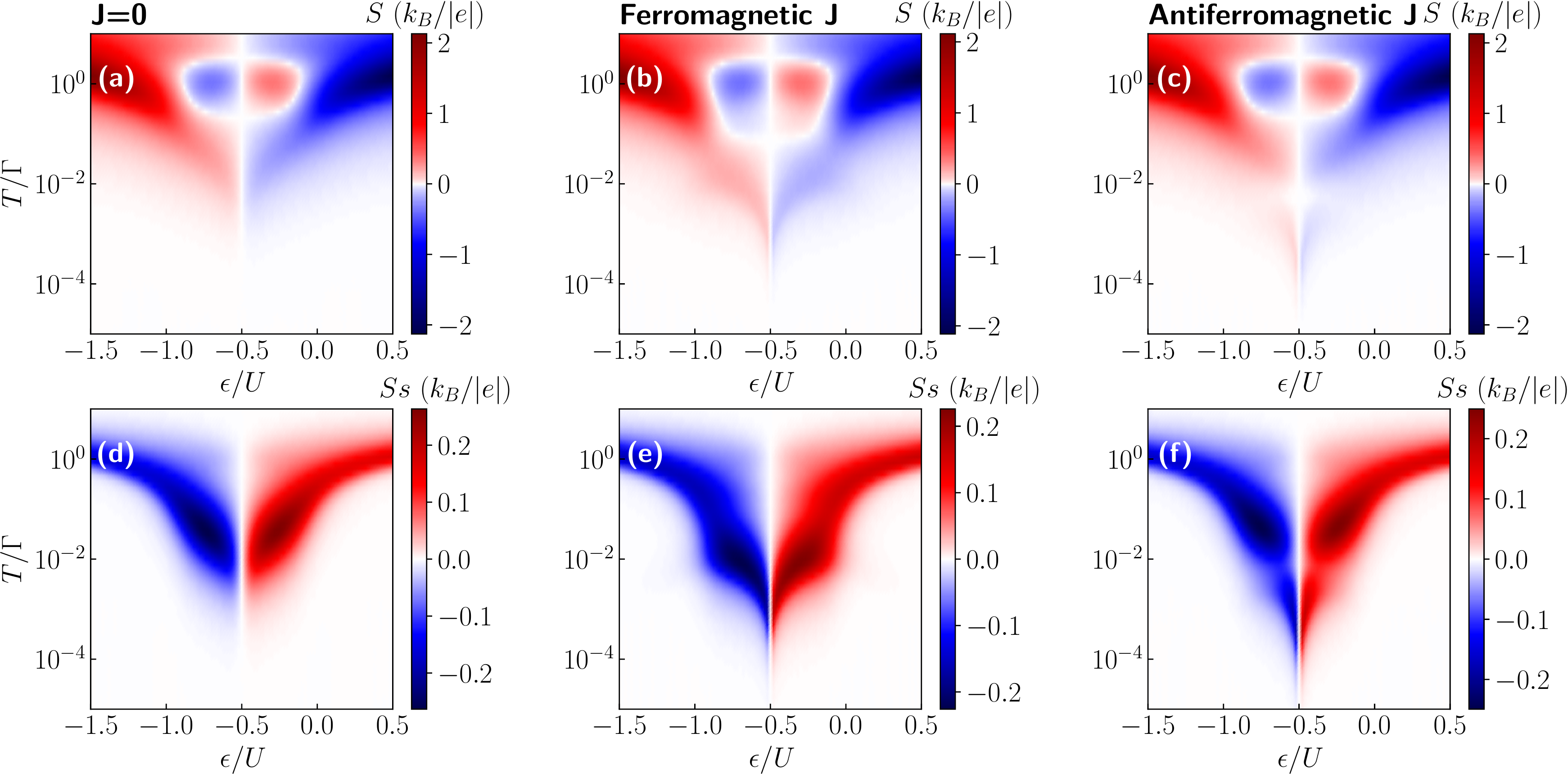}
	\caption{\label{Fig:SS2D}
		{\bf The spin Seebeck coefficient.}
		(a-c) The Seebeck and (d-f) spin Seebeck coefficient
		as a function of temperature and orbital level position
		calculated for (a,d) $J=0$
		(b,e) ferromagnetic $J$ and
		(c,f) antiferromagnetic $J$ in the case of finite spin accumulation in the leads.
		The other parameters are the same as in \fig{Fig:2DFM}.
	}
\end{figure}

The Seebeck and spin Seebeck coefficients
as a function of temperature and the position of the molecule's orbital level
calculated for different values of exchange interaction $J$ are shown in \fig{Fig:SS2D}. 
First of all, we note that the Seebeck effect at higher temperatures $T\approx \Gamma$
behaves generally very similarly as in the case of no spin accumulation in the leads,
however, its low-temperature behavior is changed. 
While in the absence of spin accumulation, $S$ exhibits an additional
sign change for $-U\lesssim \e \lesssim 0$ and $T\lesssim 0.01 \Gamma$,
in the case of long spin relaxation these features are not seen anymore.
Instead, the Seebeck coefficient changes sign only once around
$T\approx \Gamma$ and extends to low temperatures till 
it is quenched by the exchange field, see the first row of \fig{Fig:SS2D}. 
On the other hand, the spin Seebeck effect displays completely different behavior.
As can be seen, the only sign change occurs across
the particle-hole symmetry point when tuning the position of the
molecule's orbital level. Moreover, the overall behavior
is reversed as compared to the Seebeck coefficient.
In the regions where $S$ is generally negative
$S_S$ is positive and vice versa. This is directly
associated with the definition of $S$ and $S_S$---while
$S$ captures the spin-resolved contributions due to 
hole and electron processes, $S_S$ presents mainly 
the difference between the spin-resolved contributions.
As can be seen in the bottom row of \fig{Fig:SS2D},
the spin-up contribution dominates the thermopower 
for $\e>-U/2$, while for $\e<-U/2$ the spin-down
thermopower is dominant. Such behavior is visible
in all considered cases, namely, for $J=0$ as well as
for ferromagnetic and antiferromagnetic exchange interaction.
However, there are some subtle differences.
First of all, the spin Seebeck effect is finite 
for lower temperatures in the case of finite $J$,
as compared to the case with $J=0$.
Moreover, while for ferromagnetic $J$,
$S_S$ exhibits a local maximum for $T\approx 0.01 \Gamma$,
for antiferromagnetic $J$, there is a small local minimum 
in the temperature dependence of spin Seebeck effect,
see \fig{Fig:SS2D}(f).

\begin{figure}[t]
	\centering
	\includegraphics[width=0.8\columnwidth]{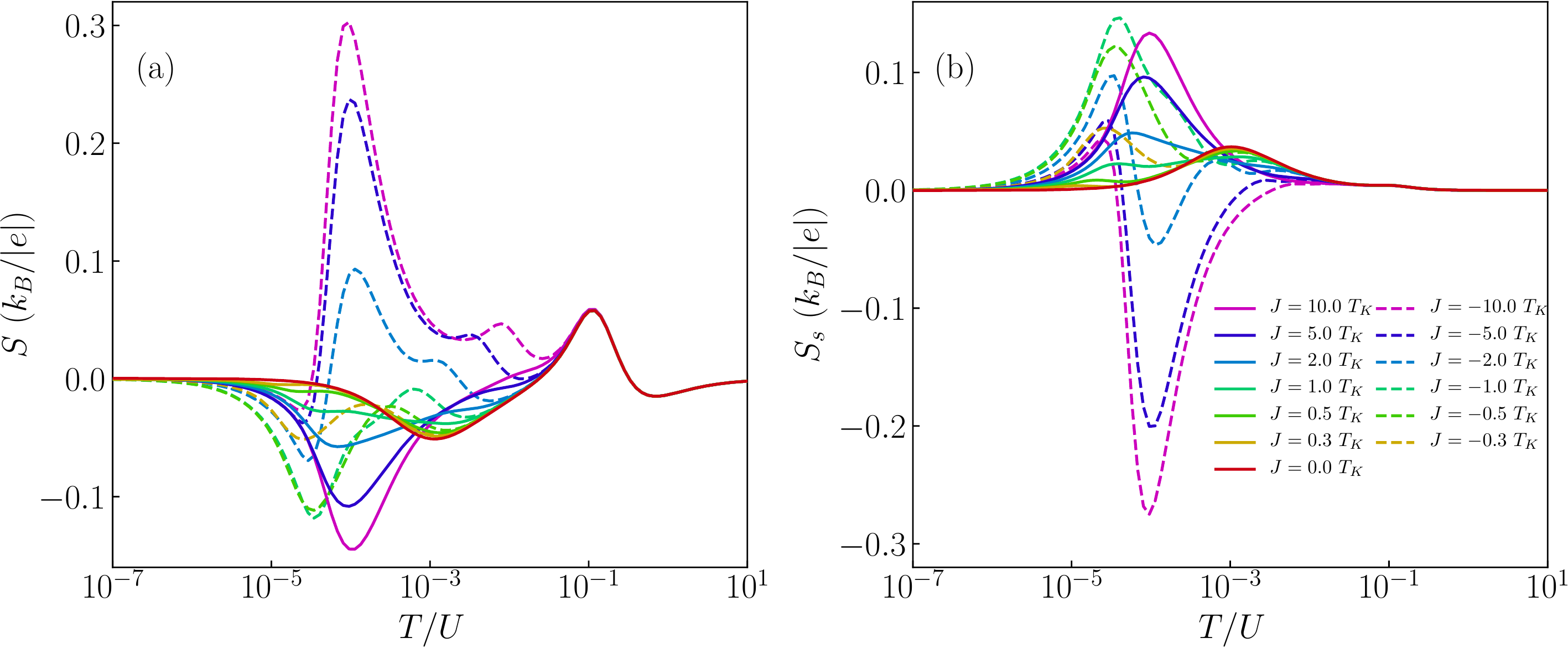}
	\caption{\label{Fig:SS1D}
		{\bf  Spin thermopower for different values of exchange interaction.}
		(a) The Seebeck and (b) spin Seebeck effect as a function of temperature
		for different values of exchange interaction $J$, as indicated.
		The parameters are the same as in \fig{Fig:2DFM} with $\e=-0.48U$.
	}
\end{figure}

The above-mentioned features are better visible in 
the temperature dependence of the thermopower and spin thermopower 
presented in \fig{Fig:SS1D}. Moreover, it turns out
that the above discussion is not fully complete,
since one can now clearly observe that a sign change of both the Seebeck
as well as spin Seebeck effect can develop,
provided that the exchange interaction is sufficiently large.
In the case of ferromagnetic exchange, a large minimum in $S$
develops with increasing $J$ at energy scale corresponding to the exchange field,
see \fig{Fig:SS1D}(a). On the other hand, for the spin Seebeck effect
a maximum forms at the same temperature for which $S$ exhibits a minimum.
An interesting behavior occurs for antiferromagnetic exchange interaction,
when with increasing $|J|$, both the thermopower and spin thermopower change
sign once $J\lesssim -T_K$. More specifically, $S$ exhibits then a pronounced maximum,
whereas a considerable minimum develops in $S_S$ for $T\approx \exch$, see \fig{Fig:SS1D}.
The sign change of the thermopower is associated with the fact that the exchange field effects,
which determine the sign of $S$ and $S_S$ for low values of $J$,
become overwhelmed by antiferromagnetic $J$,
once the exchange interaction becomes sufficiently large, i.e. $|J|\gtrsim \exch$.

\begin{figure}[t]
	\centering
	\includegraphics[width=0.8\columnwidth]{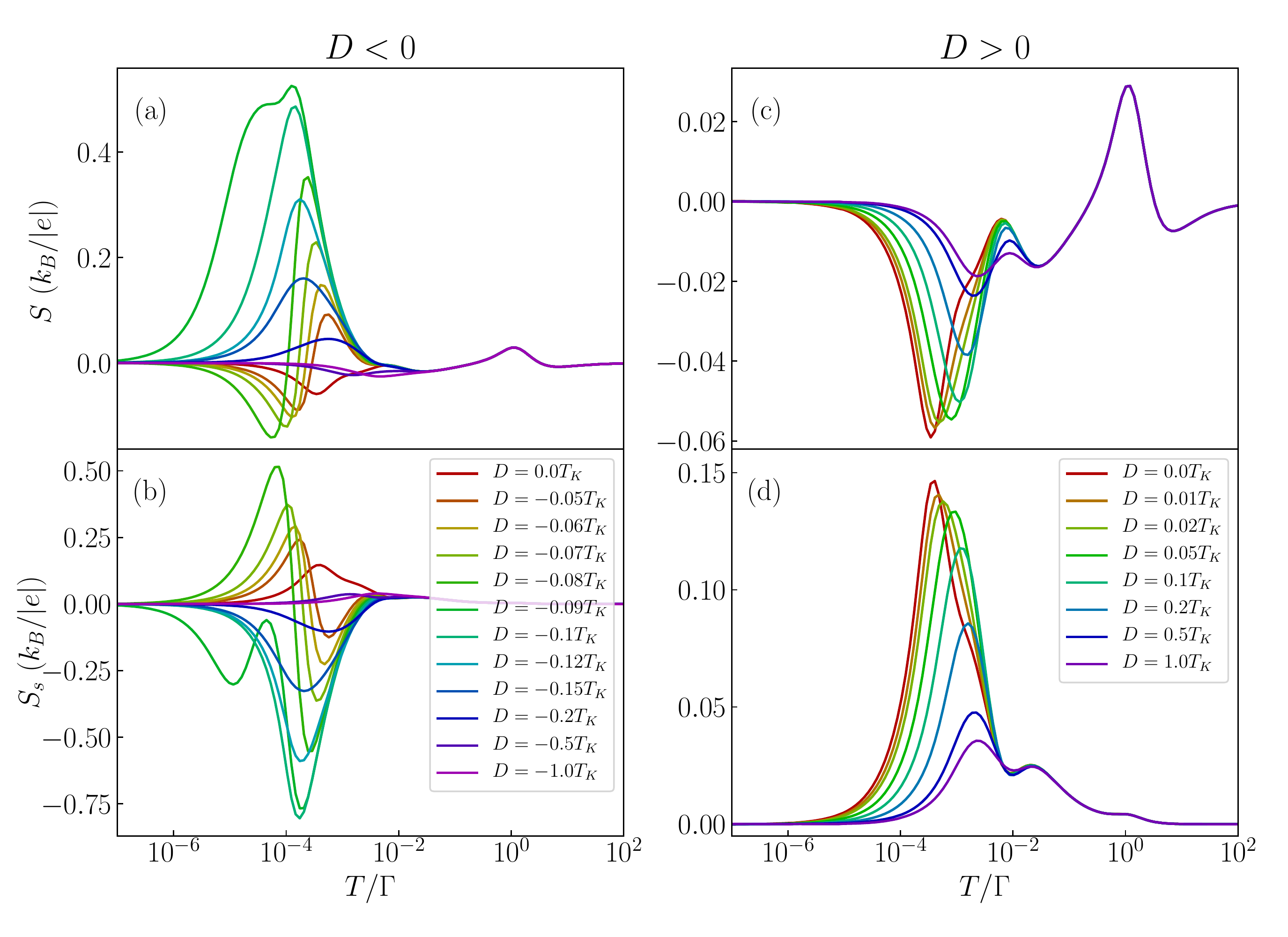}
	\caption{\label{Fig:SS1DD}
		{\bf  Spin thermopower for different values of magnetic anisotropy.}
		The temperature dependence of (a,c) the thermopower and (b,d) the spin thermopower
		in the case of antiferromagnetic exchange interaction $J=-T_K$ for selected values of magnetic anisotropy $D$.
		The left (right) column presents the case of easy-plane (easy-axis) type of magnetic anisotropy.
		The parameters are the same as in \fig{Fig:SS1D}.
	}
\end{figure}

To further understand the spin thermopower of magnetic molecules,
we present the impact of magnetic anisotropy on the Seebeck and
spin Seebeck effects for the case of antiferromagnetic exchange interaction in \fig{Fig:SS1DD}.
First of all, for the case of uniaxial type of magnetic anisotropy,
as shown in the right column of \fig{Fig:SS1DD},
there is a suppression in the minima (maxima) present in the Seebeck (spin Seebeck)
effect with increasing anisotropy. This behavior is in fact
similar to that observed in the case of no spin accumulation discussed in previous section.
However, the case of an easy plane type of anisotropy presented in the left column of \fig{Fig:SS1DD}
shows far more interesting behavior. The thermopower and spin thermopower
exhibit a change in sign around $T\approx \exch \approx 10^{-4}$
for values of $D\lesssim -0.1 T_K$.
For temperatures $T\lesssim \exch$, a considerable peak, 
positive (negative) for the spin Seebeck (Seebeck) effect,
is formed. On the other hand, for $T\gtrsim \exch$
an additional maximum (minimum) develops for $S$ ($S_S$),
see the left column of \fig{Fig:SS1DD}.
Further increase in the magnitude of anisotropy $D$ suppresses the local extrema
in the thermopower, as seen in the case of no spin accumulation.
Consequently, one observes a strong nonmonotonic dependence
of the (spin) Seebeck effect on the value of the easy-plane anisotropy.
Large values of $S$ and $S_S$ develop for such intrinsic parameters of the 
molecule that the temperature dependence of $G$ is most spectacular,
i.e. when the Kondo effect starts being restored with increasing $|D|$.
However, once a plateau in $G$ is formed at low temperatures due to the Kondo effect,
which happens for considerable values of $D<0$,
the thermopower becomes suppressed, see \fig{Fig:SS1DD}.

\section*{Discussion}

We have determined the thermoelectric properties of
large-spin magnetic molecules attached to both
nonmagnetic and ferromagnetic electrodes. 
Our analysis was focused on the strong correlation regime
where the Kondo effect can emerge. To accurately 
address the system's behavior in this nonperturbative regime, we
have employed the numerical renormalization group method,
which allowed us to study the electrical and heat conductances,
the Seebeck effect and the figure of merit in the full parameter space
of considered molecular junction. In particular, we have
considered the cases of both ferromagnetic and antiferromagnetic
exchange interaction $J$ between the orbital level of the molecule
and its magnetic core. Moreover, we have also analyzed the effect 
of finite magnetic anisotropy of the molecule. 

In the case of nonmagnetic contacts, we have shown that
the behavior of the Seebeck effect strongly depends on the type of
exchange interaction $J$. In the case of ferromagnetic
exchange, we have shown that the Seebeck coefficient
displays features qualitatively similar to the case of $J=0$ (quantum dot case).
However, due to a reduced Kondo temperature, the sign change
in $S$ occurs at a lower temperature and the thermopower
was found to retain finite values extending to much lower $T$
compared to the quantum dot case. Interestingly, in the case 
of antiferromagnetic exchange interaction, we have found 
a new sign change of the thermopower, at energy scale corresponding
to the exchange interaction between the orbital level and molecule's core spin.
Moreover, we have also determined the influence of finite
magnetic anisotropy on the thermoelectric properties of magnetic molecule.
Finite magnetic anisotropy gives rise to new qualitative features especially in the case of antiferromagnetic $J$,
where a new sign change of thermopower occurs
in the case of easy-plane magnetic anisotropy. 

On the other hand, when the leads are ferromagnetic,
depending on the spin relaxation time, spin accumulation may be generated
in the contacts giving rise to the spin Seebeck effect. First, we have focused on
the impact of the spin-resolved tunneling on the thermopower in the absence
of spin accumulation. We have shown that in the case of ferromagnetic 
exchange interaction there is an additional sign change in the temperature
dependence of the thermopower. On the other hand, the low-temperature
Seebeck effect has been found to strongly depend on the position
of the molecule's orbital level, which conditioned the strength
of the exchange field in the molecule. Generally, the Seebeck 
effect becomes quenched once the temperature gets smaller
than the corresponding exchange field.

Finally, we have assumed that the spin relaxation time in the leads is long,
so that a spin bias can be generated in the system, resulting in the spin Seebeck effect.
We have shown that for relatively low values of molecule's exchange interaction
the spin Seebeck effect changes sign only when tuning
the orbital level across the particle-hole symmetry point.
This is associated with the fact, that the sign of the spin thermopower
is conditioned by the sign of the exchange field, which
changes only when crossing $\e=-U/2$.
However, if the molecule's antiferromagnetic exchange interaction
becomes larger than the Kondo temperature,
the temperature dependence of the spin Seebeck effect
can exhibit a sign change. This feature is associated
with the interplay of exchange field, antiferromagnetic interaction between
the orbital level and molecule's core spin and the Kondo correlations.
A similarly nontrivial behavior have been observed in the case of 
finite easy-plane magnetic anisotropy. We have found that 
the spin Seebeck effect exhibits then a nonmonotonic dependence on the
magnitude of anisotropy. This effect is related to a revival of the Kondo effect
when the anisotropy becomes large enough to bring about the two-fold degenerate
ground state of the molecule, i.e. when the easy-plane anisotropy wins over both the
exchange field as well as the molecule's antiferromagnetic exchange interaction.

\new{
As far as the experimental progress is concerned,
the field of molecular spin caloritronics 
is rather at its initial stage of development. The model studied
here can however also describe quantum dots or impurities coupled
to a large spin. It is therefore worth mentioning that, recently, there have been successful
measurements of thermopower of Kondo-correlated quantum dots
\cite{Scheibner2005Oct,Svilans2018Nov,Dutta2019Jan}. Moreover,
spin-resolved electronic transport properties of molecular junctions have already been extensively
studied \cite{Pasupathy_Science306/2004,Hauptmann_NaturePhys.4/2008,
	Gaass_Phys.Rev.Lett.107/2011,bordoloi_double_2020}.
Therefore, it seems that all the necessary ingredients are at hand
and, with the state-of-the-art apparatus, it should be possible to explore the effects presented in this paper.
We do hope that our work will foster further experimental efforts in this direction.
}


\subsection*{Methods}
\label{subsec:method}

The thermoelectric properties of the system in the  linear response regime
can be characterized by the Onsager integrals, $L_{n\s}$,
which depend on the spin-resolved transmission coefficient $\T_{\s}(\omega)$.
The transmission coefficient $\T_{\s}(\omega)$ can be related to the spin-dependent spectral function $A_{\s}(\omega)$ using the relation
\be
\T_{\s}(\w) = \frac{4\Gamma_{L\s}\Gamma_{R\s}}{\Gamma_{L\s} + \Gamma_{R\s}}\pi A_{\s}(\w) ,
\ee
where $A_{\s}(\w) = -Im[G_{\s}^R(\w)]/\pi$ and $G_{\s}^R(\w)$
is the Fourier transform of the retarded Green's function of the molecule's orbital level,
$G_{\s}^{R}(t)=-i\theta(t)\langle\{\hat{d}_\s(t),\hat{d}_\s^\dagger(0)\}\rangle$.
The main task is thus to accurately determine the spectral function of the system.
One of the most powerful methods in this regard
is the Wilson's numerical renormalization group method \cite{Wilson_Rev.Mod.Phys.47/1975,Bulla_Rev.Mod.Phys.80/2008,Legeza2008Sep,Toth2008Dec},
which allows for nonperturbative treatment of all correlations  in the system.
In this method one performs a discretization of the conduction band
with discretization parameter $\Lambda$ and, consecutively, a tridiagonalization
of the Hamiltonian describing such discretized system is performed.
Eventually, one obtains the following NRG Hamiltonian
\cite{Wilson_Rev.Mod.Phys.47/1975}
\be
\hat{H}_{\rm NRG} = \hat{H}_{\text{mol}}
+ \sum_\sigma\sqrt{\frac{2W\Gamma_\sigma}{\pi}} \left(\hat{d}_{\sigma}^\dagger \hat{f}_{0\sigma} + \hat{f}_{0\sigma}^\dag \hat{d}_\sigma\right)
+ \sum_\sigma \sum_{n=0} \xi_n \left(\hat{f}_{n\sigma}^\dag \hat{f}_{n+1\sigma} + \hat{f}_{n+1\sigma}^\dag \hat{f}_{n\sigma}\right),
\ee
in which the molecule is coupled to the first site of
the tight-binding chain with exponentially decaying hoppings $\xi_n$.
Here, $\hat{f}_{n\sigma}^\dag$ creates a spin-$\s$ electron
at the Wilson site $n$ and $W$ denotes the band halfwidth, which is used as energy unit $W\equiv 1$.
This Hamiltonian is then solved in an iterative fashion
by retaining a fixed number of states $N_K$.
While the kept states are used to construct the statespace
for the next iteration, the states that are discarded during calculation
play a vital role in the problem as these states are used to construct
the complete many-body basis of the full NRG Hamiltonian
\cite{Anders2005}.
The discarded states are then used for the calculation
of quantities of interest with the aid of the full density matrix \cite{Weichselbaum_Phys.Rev.Lett.99/2007}.
In our calculations we have exploited the Abelian symmetries
for the system's spin $z$th component and charge. We have used $\Lambda = 2$
and kept at least $N_K=2000$ states in the iteration. Moreover, we have determined the 
Onsager transport coefficients from the raw NRG data
without the need of broadening the Dirac delta peaks \cite{Weymann2013Aug}.

\section*{Acknowledgements}

This work was supported by the National Science Centre
in Poland through the Project No. 2017/27/B/ST3/00621.

\section*{Author contributions statement}

A.M. performed the calculations and prepared the figures.
Both authors contributed to the manuscript preparation and
results interpretation. I.W. coordinated the whole project.

\section*{Additional information}
The authors declare no competing financial interests.


\begin{thebibliography}{10}
\expandafter\ifx\csname url\endcsname\relax
  \def\url#1{\texttt{#1}}\fi
\expandafter\ifx\csname urlprefix\endcsname\relax\def\urlprefix{URL }\fi
\providecommand{\bibinfo}[2]{#2}
\providecommand{\eprint}[2][]{\url{#2}}

\bibitem{Giazotto2006Mar}
\bibinfo{author}{Giazotto, F.},
  \bibinfo{author}{Heikkil{\ifmmode\ddot{a}\else\"{a}\fi}, T.~T.},
  \bibinfo{author}{Luukanen, A.}, \bibinfo{author}{Savin, A.~M.} \&
  \bibinfo{author}{Pekola, J.~P.}
\newblock \bibinfo{title}{{Opportunities for mesoscopics in thermometry and
  refrigeration: Physics and applications}}.
\newblock \emph{\bibinfo{journal}{Rev. Mod. Phys.}}
  \textbf{\bibinfo{volume}{78}}, \bibinfo{pages}{217--274}
  (\bibinfo{year}{2006}).

\bibitem{Szczech2011Mar}
\bibinfo{author}{Szczech, J.~R.}, \bibinfo{author}{Higgins, J.~M.} \&
  \bibinfo{author}{Jin, S.}
\newblock \bibinfo{title}{{Enhancement of the thermoelectric properties in
  nanoscale and nanostructured materials}}.
\newblock \emph{\bibinfo{journal}{J. Mater. Chem.}}
  \textbf{\bibinfo{volume}{21}}, \bibinfo{pages}{4037--4055}
  (\bibinfo{year}{2011}).

\bibitem{Heremans2013Jul}
\bibinfo{author}{Heremans, J.~P.}, \bibinfo{author}{Dresselhaus, M.~S.},
  \bibinfo{author}{Bell, L.~E.} \& \bibinfo{author}{Morelli, D.~T.}
\newblock \bibinfo{title}{{When thermoelectrics reached the nanoscale}}.
\newblock \emph{\bibinfo{journal}{Nat. Nanotechnol.}}
  \textbf{\bibinfo{volume}{8}}, \bibinfo{pages}{471--473}
  (\bibinfo{year}{2013}).

\bibitem{Sanchez2014Nov}
\bibinfo{author}{S{\ifmmode\acute{a}\else\'{a}\fi}nchez, D.} \&
  \bibinfo{author}{Linke, H.}
\newblock \bibinfo{title}{{Focus on thermoelectric effects in nanostructures}}.
\newblock \emph{\bibinfo{journal}{New J. Phys.}} \textbf{\bibinfo{volume}{16}},
  \bibinfo{pages}{110201} (\bibinfo{year}{2014}).

\bibitem{Benenti2017Jun}
\bibinfo{author}{Benenti, G.}, \bibinfo{author}{Casati, G.},
  \bibinfo{author}{Saito, K.} \& \bibinfo{author}{Whitney, R.~S.}
\newblock \bibinfo{title}{{Fundamental aspects of steady-state conversion of
  heat to work at the nanoscale}}.
\newblock \emph{\bibinfo{journal}{Phys. Rep.}} \textbf{\bibinfo{volume}{694}},
  \bibinfo{pages}{1--124} (\bibinfo{year}{2017}).

\bibitem{Hicks1993May}
\bibinfo{author}{Hicks, L.~D.} \& \bibinfo{author}{Dresselhaus, M.~S.}
\newblock \bibinfo{title}{{Effect of quantum-well structures on the
  thermoelectric figure of merit}}.
\newblock \emph{\bibinfo{journal}{Phys. Rev. B}} \textbf{\bibinfo{volume}{47}},
  \bibinfo{pages}{12727--12731} (\bibinfo{year}{1993}).

\bibitem{Hicks1993Jun}
\bibinfo{author}{Hicks, L.~D.} \& \bibinfo{author}{Dresselhaus, M.~S.}
\newblock \bibinfo{title}{{Thermoelectric figure of merit of a one-dimensional
  conductor}}.
\newblock \emph{\bibinfo{journal}{Phys. Rev. B}} \textbf{\bibinfo{volume}{47}},
  \bibinfo{pages}{16631--16634(R)} (\bibinfo{year}{1993}).

\bibitem{Mahan1996Jul}
\bibinfo{author}{Mahan, G.~D.} \& \bibinfo{author}{Sofo, J.~O.}
\newblock \bibinfo{title}{{The best thermoelectric}}.
\newblock \emph{\bibinfo{journal}{Proc. Natl. Acad. Sci. U.S.A.}}
  \textbf{\bibinfo{volume}{93}}, \bibinfo{pages}{7436--7439}
  (\bibinfo{year}{1996}).

\bibitem{Beenakker1992Oct}
\bibinfo{author}{Beenakker, C. W.~J.} \& \bibinfo{author}{Staring, A. A.~M.}
\newblock \bibinfo{title}{{Theory of the thermopower of a quantum dot}}.
\newblock \emph{\bibinfo{journal}{Phys. Rev. B}} \textbf{\bibinfo{volume}{46}},
  \bibinfo{pages}{9667--9676} (\bibinfo{year}{1992}).

\bibitem{Scheibner2005Oct}
\bibinfo{author}{Scheibner, R.}, \bibinfo{author}{Buhmann, H.},
  \bibinfo{author}{Reuter, D.}, \bibinfo{author}{Kiselev, M.~N.} \&
  \bibinfo{author}{Molenkamp, L.~W.}
\newblock \bibinfo{title}{{Thermopower of a Kondo Spin-Correlated Quantum
  Dot}}.
\newblock \emph{\bibinfo{journal}{Phys. Rev. Lett.}}
  \textbf{\bibinfo{volume}{95}}, \bibinfo{pages}{176602}
  (\bibinfo{year}{2005}).

\bibitem{Reddy2007Mar}
\bibinfo{author}{Reddy, P.}, \bibinfo{author}{Jang, S.-Y.},
  \bibinfo{author}{Segalman, R.~A.} \& \bibinfo{author}{Majumdar, A.}
\newblock \bibinfo{title}{{Thermoelectricity in Molecular Junctions}}.
\newblock \emph{\bibinfo{journal}{Science}} \textbf{\bibinfo{volume}{315}},
  \bibinfo{pages}{1568--1571} (\bibinfo{year}{2007}).

\bibitem{Dubi2011Mar}
\bibinfo{author}{Dubi, Y.} \& \bibinfo{author}{Di~Ventra, M.}
\newblock \bibinfo{title}{{Colloquium: Heat flow and thermoelectricity in
  atomic and molecular junctions}}.
\newblock \emph{\bibinfo{journal}{Rev. Mod. Phys.}}
  \textbf{\bibinfo{volume}{83}}, \bibinfo{pages}{131--155}
  (\bibinfo{year}{2011}).

\bibitem{Trocha2012Feb}
\bibinfo{author}{Trocha, P.} \&
  \bibinfo{author}{Barna{\ifmmode\acute{s}\else\'{s}\fi}, J.}
\newblock \bibinfo{title}{{Large enhancement of thermoelectric effects in a
  double quantum dot system due to interference and Coulomb correlation
  phenomena}}.
\newblock \emph{\bibinfo{journal}{Phys. Rev. B}} \textbf{\bibinfo{volume}{85}},
  \bibinfo{pages}{085408} (\bibinfo{year}{2012}).

\bibitem{Thierschmann2016Dec}
\bibinfo{author}{Thierschmann, H.},
  \bibinfo{author}{S{\ifmmode\acute{a}\else\'{a}\fi}nchez, R.},
  \bibinfo{author}{Sothmann, B.}, \bibinfo{author}{Buhmann, H.} \&
  \bibinfo{author}{Molenkamp, L.~W.}
\newblock \bibinfo{title}{{Thermoelectrics with Coulomb-coupled quantum dots}}.
\newblock \emph{\bibinfo{journal}{C. R. Phys.}} \textbf{\bibinfo{volume}{17}},
  \bibinfo{pages}{1109--1122} (\bibinfo{year}{2016}).

\bibitem{Jaliel2019Sep}
\bibinfo{author}{Jaliel, G.} \emph{et~al.}
\newblock \bibinfo{title}{{Experimental Realization of a Quantum Dot Energy
  Harvester}}.
\newblock \emph{\bibinfo{journal}{Phys. Rev. Lett.}}
  \textbf{\bibinfo{volume}{123}}, \bibinfo{pages}{117701}
  (\bibinfo{year}{2019}).

\bibitem{Kleeorin2019Dec}
\bibinfo{author}{Kleeorin, Y.} \emph{et~al.}
\newblock \bibinfo{title}{{How to measure the entropy of a mesoscopic system
  via thermoelectric transport}}.
\newblock \emph{\bibinfo{journal}{Nat. Commun.}} \textbf{\bibinfo{volume}{10}},
  \bibinfo{pages}{1--8} (\bibinfo{year}{2019}).

\bibitem{Kondo_Prog.Theor.Phys32/1964}
\bibinfo{author}{Kondo, J.}
\newblock \bibinfo{title}{{Resistance minimum in dilute magnetic alloys}}.
\newblock \emph{\bibinfo{journal}{Prog. Theor. Phys}}
  \textbf{\bibinfo{volume}{32}}, \bibinfo{pages}{37} (\bibinfo{year}{1964}).

\bibitem{Hewson_book}
\bibinfo{author}{Hewson, A.~C.}
\newblock \emph{\bibinfo{title}{{The Kondo problem to heavy fermions}}}
  (\bibinfo{publisher}{Cambridge University Press},
  \bibinfo{address}{Cambridge}, \bibinfo{year}{1997}).

\bibitem{Goldhaber_Nature391/98}
\bibinfo{author}{Goldhaber-Gordon, D.} \emph{et~al.}
\newblock \bibinfo{title}{{The Kondo effect in a single-electron transistor}}.
\newblock \emph{\bibinfo{journal}{Nature (London)}}
  \textbf{\bibinfo{volume}{391}}, \bibinfo{pages}{156--159}
  (\bibinfo{year}{1998}).

\bibitem{Cronenwett_Science281/1998}
\bibinfo{author}{Cronenwett, S.}, \bibinfo{author}{Oosterkamp, T.} \&
  \bibinfo{author}{Kouwenhoven, L.}
\newblock \bibinfo{title}{{A tunable Kondo effect in quantum dots}}.
\newblock \emph{\bibinfo{journal}{Science}} \textbf{\bibinfo{volume}{281}},
  \bibinfo{pages}{540} (\bibinfo{year}{1998}).

\bibitem{Costi2010Jun}
\bibinfo{author}{Costi, T.~A.} \&
  \bibinfo{author}{Zlati{\ifmmode\acute{c}\else\'{c}\fi}, V.}
\newblock \bibinfo{title}{{Thermoelectric transport through strongly correlated
  quantum dots}}.
\newblock \emph{\bibinfo{journal}{Phys. Rev. B}} \textbf{\bibinfo{volume}{81}},
  \bibinfo{pages}{235127} (\bibinfo{year}{2010}).

\bibitem{Tooski2014May}
\bibinfo{author}{Tooski, S.~B.},
  \bibinfo{author}{Ram{\ifmmode\check{s}\else\v{s}\fi}ak, A.},
  \bibinfo{author}{Bu{\l}ka, B.~R.} \&
  \bibinfo{author}{{\ifmmode\check{Z}\else\v{Z}\fi}itko, R.}
\newblock \bibinfo{title}{{Effect of assisted hopping on thermopower in an
  interacting quantum dot}}.
\newblock \emph{\bibinfo{journal}{New J. Phys.}} \textbf{\bibinfo{volume}{16}},
  \bibinfo{pages}{055001} (\bibinfo{year}{2014}).

\bibitem{Wojcik2016Feb}
\bibinfo{author}{W{\ifmmode\acute{o}\else\'{o}\fi}jcik, K.~P.} \&
  \bibinfo{author}{Weymann, I.}
\newblock \bibinfo{title}{{Thermopower of strongly correlated T-shaped double
  quantum dots}}.
\newblock \emph{\bibinfo{journal}{Phys. Rev. B}} \textbf{\bibinfo{volume}{93}},
  \bibinfo{pages}{085428} (\bibinfo{year}{2016}).

\bibitem{Nguyen2020Jul}
\bibinfo{author}{Nguyen, T. K.~T.} \& \bibinfo{author}{Kiselev, M.~N.}
\newblock \bibinfo{title}{{Thermoelectric Transport in a Three-Channel Charge
  Kondo Circuit}}.
\newblock \emph{\bibinfo{journal}{Phys. Rev. Lett.}}
  \textbf{\bibinfo{volume}{125}}, \bibinfo{pages}{026801}
  (\bibinfo{year}{2020}).

\bibitem{Svilans2018Nov}
\bibinfo{author}{Svilans, A.} \emph{et~al.}
\newblock \bibinfo{title}{{Thermoelectric Characterization of the Kondo
  Resonance in Nanowire Quantum Dots}}.
\newblock \emph{\bibinfo{journal}{Phys. Rev. Lett.}}
  \textbf{\bibinfo{volume}{121}}, \bibinfo{pages}{206801}
  (\bibinfo{year}{2018}).

\bibitem{Dutta2019Jan}
\bibinfo{author}{Dutta, B.} \emph{et~al.}
\newblock \bibinfo{title}{{Direct Probe of the Seebeck Coefficient in a
  Kondo-Correlated Single-Quantum-Dot Transistor}}.
\newblock \emph{\bibinfo{journal}{Nano Lett.}} \textbf{\bibinfo{volume}{19}},
  \bibinfo{pages}{506--511} (\bibinfo{year}{2019}).

\bibitem{Uchida2008Oct}
\bibinfo{author}{Uchida, K.} \emph{et~al.}
\newblock \bibinfo{title}{{Observation of the spin Seebeck effect}}.
\newblock \emph{\bibinfo{journal}{Nature}} \textbf{\bibinfo{volume}{455}},
  \bibinfo{pages}{778--781} (\bibinfo{year}{2008}).

\bibitem{Johnson2010Mar}
\bibinfo{author}{Johnson, M.}
\newblock \bibinfo{title}{{Spin caloritronics and the thermomagnetoelectric
  system}}.
\newblock \emph{\bibinfo{journal}{Solid State Commun.}}
  \textbf{\bibinfo{volume}{150}}, \bibinfo{pages}{543--547}
  (\bibinfo{year}{2010}).

\bibitem{Bauer2012May}
\bibinfo{author}{Bauer, G. E.~W.}, \bibinfo{author}{Saitoh, E.} \&
  \bibinfo{author}{van Wees, B.~J.}
\newblock \bibinfo{title}{{Spin caloritronics}}.
\newblock \emph{\bibinfo{journal}{Nat. Mater.}} \textbf{\bibinfo{volume}{11}},
  \bibinfo{pages}{391--399} (\bibinfo{year}{2012}).

\bibitem{Yu2017Mar}
\bibinfo{author}{Yu, H.}, \bibinfo{author}{Brechet, S.~D.} \&
  \bibinfo{author}{Ansermet, J.-P.}
\newblock \bibinfo{title}{{Spin caloritronics, origin and outlook}}.
\newblock \emph{\bibinfo{journal}{Phys. Lett. A}}
  \textbf{\bibinfo{volume}{381}}, \bibinfo{pages}{825--837}
  (\bibinfo{year}{2017}).

\bibitem{Back2019Mar}
\bibinfo{author}{Back, C.~H.}, \bibinfo{author}{Bauer, G. E.~W.} \&
  \bibinfo{author}{Zink, B.~L.}
\newblock \bibinfo{title}{{Special issue on spin caloritronics}}.
\newblock \emph{\bibinfo{journal}{J. Phys. D: Appl. Phys.}}
  \textbf{\bibinfo{volume}{52}}, \bibinfo{pages}{230301}
  (\bibinfo{year}{2019}).

\bibitem{uchida_transport_2021}
\bibinfo{author}{Uchida, K.-i.}
\newblock \bibinfo{title}{Transport phenomena in spin caloritronics}.
\newblock \emph{\bibinfo{journal}{Proceedings of the Japan Academy, Series B}}
  \textbf{\bibinfo{volume}{97}}, \bibinfo{pages}{69--88}
  (\bibinfo{year}{2021}).

\bibitem{Krawiec2006Feb}
\bibinfo{author}{Krawiec, M.} \&
  \bibinfo{author}{Wysoki{\ifmmode\acute{n}\else\'{n}\fi}ski, K.~I.}
\newblock \bibinfo{title}{{Thermoelectric effects in strongly interacting
  quantum dot coupled to ferromagnetic leads}}.
\newblock \emph{\bibinfo{journal}{Phys. Rev. B}} \textbf{\bibinfo{volume}{73}},
  \bibinfo{pages}{075307} (\bibinfo{year}{2006}).

\bibitem{Dubi2009Feb}
\bibinfo{author}{Dubi, Y.} \& \bibinfo{author}{Di~Ventra, M.}
\newblock \bibinfo{title}{{Thermospin effects in a quantum dot connected to
  ferromagnetic leads}}.
\newblock \emph{\bibinfo{journal}{Phys. Rev. B}} \textbf{\bibinfo{volume}{79}},
  \bibinfo{pages}{081302} (\bibinfo{year}{2009}).

\bibitem{Swirkowicz2009Nov}
\bibinfo{author}{{\ifmmode\acute{S}\else\'{S}\fi}wirkowicz, R.},
  \bibinfo{author}{Wierzbicki, M.} \&
  \bibinfo{author}{Barna{\ifmmode\acute{s}\else\'{s}\fi}, J.}
\newblock \bibinfo{title}{{Thermoelectric effects in transport through quantum
  dots attached to ferromagnetic leads with noncollinear magnetic moments}}.
\newblock \emph{\bibinfo{journal}{Phys. Rev. B}} \textbf{\bibinfo{volume}{80}},
  \bibinfo{pages}{195409} (\bibinfo{year}{2009}).

\bibitem{Weymann2013Aug}
\bibinfo{author}{Weymann, I.} \&
  \bibinfo{author}{Barna{\ifmmode\acute{s}\else\'{s}\fi}, J.}
\newblock \bibinfo{title}{{Spin thermoelectric effects in Kondo quantum dots
  coupled to ferromagnetic leads}}.
\newblock \emph{\bibinfo{journal}{Phys. Rev. B}} \textbf{\bibinfo{volume}{88}},
  \bibinfo{pages}{085313} (\bibinfo{year}{2013}).

\bibitem{Weymann2016Jan}
\bibinfo{author}{Weymann, I.}
\newblock \bibinfo{title}{{Boosting spin-caloritronic effects by attractive
  correlations in molecular junctions}}.
\newblock \emph{\bibinfo{journal}{Sci. Rep.}} \textbf{\bibinfo{volume}{6}},
  \bibinfo{pages}{1--10} (\bibinfo{year}{2016}).

\bibitem{Rejec2012Feb}
\bibinfo{author}{Rejec, T.},
  \bibinfo{author}{{\ifmmode\check{Z}\else\v{Z}\fi}itko, R.},
  \bibinfo{author}{Mravlje, J.} \&
  \bibinfo{author}{Ram{\ifmmode\check{s}\else\v{s}\fi}ak, A.}
\newblock \bibinfo{title}{{Spin thermopower in interacting quantum dots}}.
\newblock \emph{\bibinfo{journal}{Phys. Rev. B}} \textbf{\bibinfo{volume}{85}},
  \bibinfo{pages}{085117} (\bibinfo{year}{2012}).

\bibitem{Costi2019Oct}
\bibinfo{author}{Costi, T.~A.}
\newblock \bibinfo{title}{{Magnetic field dependence of the thermopower of
  Kondo-correlated quantum dots: Comparison with experiment}}.
\newblock \emph{\bibinfo{journal}{Phys. Rev. B}}
  \textbf{\bibinfo{volume}{100}}, \bibinfo{pages}{155126}
  (\bibinfo{year}{2019}).

\bibitem{Costi2019Oct2}
\bibinfo{author}{Costi, T.~A.}
\newblock \bibinfo{title}{{Magnetic field dependence of the thermopower of
  Kondo-correlated quantum dots}}.
\newblock \emph{\bibinfo{journal}{Phys. Rev. B}}
  \textbf{\bibinfo{volume}{100}}, \bibinfo{pages}{161106}
  (\bibinfo{year}{2019}).

\bibitem{Koch_Phys.Rev.B70/2004}
\bibinfo{author}{Koch, J.}, \bibinfo{author}{von Oppen, F.},
  \bibinfo{author}{Oreg, Y.} \& \bibinfo{author}{Sela, E.}
\newblock \bibinfo{title}{{Thermopower of single-molecule devices}}.
\newblock \emph{\bibinfo{journal}{Phys. Rev. B}} \textbf{\bibinfo{volume}{70}},
  \bibinfo{pages}{195107} (\bibinfo{year}{2004}).

\bibitem{Wang2010Jul}
\bibinfo{author}{Wang, R.-Q.}, \bibinfo{author}{Sheng, L.},
  \bibinfo{author}{Shen, R.}, \bibinfo{author}{Wang, B.} \&
  \bibinfo{author}{Xing, D.~Y.}
\newblock \bibinfo{title}{{Thermoelectric Effect in Single-Molecule-Magnet
  Junctions}}.
\newblock \emph{\bibinfo{journal}{Phys. Rev. Lett.}}
  \textbf{\bibinfo{volume}{105}}, \bibinfo{pages}{057202}
  (\bibinfo{year}{2010}).

\bibitem{Misiorny2014Jun}
\bibinfo{author}{Misiorny, M.} \&
  \bibinfo{author}{Barna{\ifmmode\acute{s}\else\'{s}\fi}, J.}
\newblock \bibinfo{title}{{Spin-dependent thermoelectric effects in transport
  through a nanoscopic junction involving a spin impurity}}.
\newblock \emph{\bibinfo{journal}{Phys. Rev. B}} \textbf{\bibinfo{volume}{89}},
  \bibinfo{pages}{235438} (\bibinfo{year}{2014}).

\bibitem{Misiorny2015Apr}
\bibinfo{author}{Misiorny, M.} \&
  \bibinfo{author}{Barna{\ifmmode\acute{s}\else\'{s}\fi}, J.}
\newblock \bibinfo{title}{{Effect of magnetic anisotropy on spin-dependent
  thermoelectric effects in nanoscopic systems}}.
\newblock \emph{\bibinfo{journal}{Phys. Rev. B}} \textbf{\bibinfo{volume}{91}},
  \bibinfo{pages}{155426} (\bibinfo{year}{2015}).

\bibitem{Niu2018Nov}
\bibinfo{author}{Niu, P.}, \bibinfo{author}{Liu, L.}, \bibinfo{author}{Su, X.},
  \bibinfo{author}{Dong, L.} \& \bibinfo{author}{Luo, H.-G.}
\newblock \bibinfo{title}{{Spin current generator in a single molecular magnet
  with spin bias}}.
\newblock \emph{\bibinfo{journal}{J. Magn. Magn. Mater.}}
  \textbf{\bibinfo{volume}{465}}, \bibinfo{pages}{9--13}
  (\bibinfo{year}{2018}).

\bibitem{Hammar2019Mar}
\bibinfo{author}{Hammar, H.}, \bibinfo{author}{Vasquez~Jaramillo, J.~D.} \&
  \bibinfo{author}{Fransson, J.}
\newblock \bibinfo{title}{{Spin-dependent heat signatures of single-molecule
  spin dynamics}}.
\newblock \emph{\bibinfo{journal}{Phys. Rev. B}} \textbf{\bibinfo{volume}{99}},
  \bibinfo{pages}{115416} (\bibinfo{year}{2019}).

\bibitem{Gatteschi_Science265/1994}
\bibinfo{author}{Gatteschi, D.}, \bibinfo{author}{Caneschi, A.},
  \bibinfo{author}{Pardi, L.} \& \bibinfo{author}{Sessoli, R.}
\newblock \bibinfo{title}{{Large clusters of metal ions: the transition from
  molecular to bulk magnets}}.
\newblock \emph{\bibinfo{journal}{Science}} \textbf{\bibinfo{volume}{265}},
  \bibinfo{pages}{1054} (\bibinfo{year}{1994}).

\bibitem{Gatteschi_book}
\bibinfo{author}{Gatteschi, D.}, \bibinfo{author}{Sessoli, R.} \&
  \bibinfo{author}{Villain, J.}
\newblock \emph{\bibinfo{title}{{Molecular nanomagnets}}}
  (\bibinfo{publisher}{Oxford University Press}, \bibinfo{address}{New York},
  \bibinfo{year}{2006}).

\bibitem{Gatteschi_Angew.Chem.Int.Ed.42/2003}
\bibinfo{author}{Gatteschi, D.} \& \bibinfo{author}{Sessoli, R.}
\newblock \bibinfo{title}{Quantum tunneling of magnetization and related
  phenomena in molecular materials}.
\newblock \emph{\bibinfo{journal}{Angew. Chem. Int. Ed.}}
  \textbf{\bibinfo{volume}{42}}, \bibinfo{pages}{268--297}
  (\bibinfo{year}{2003}).

\bibitem{Wilson_Rev.Mod.Phys.47/1975}
\bibinfo{author}{Wilson, K.~G.}
\newblock \bibinfo{title}{{The renormalization group: Critical phenomena and
  the Kondo problem}}.
\newblock \emph{\bibinfo{journal}{Rev. Mod. Phys.}}
  \textbf{\bibinfo{volume}{47}}, \bibinfo{pages}{773--839}
  (\bibinfo{year}{1975}).

\bibitem{Bulla_Rev.Mod.Phys.80/2008}
\bibinfo{author}{Bulla, R.}, \bibinfo{author}{Costi, T.} \&
  \bibinfo{author}{Pruschke, T.}
\newblock \bibinfo{title}{{Numerical renormalization group method for quantum
  impurity systems}}.
\newblock \emph{\bibinfo{journal}{Rev. Mod. Phys.}}
  \textbf{\bibinfo{volume}{80}}, \bibinfo{pages}{395--450}
  (\bibinfo{year}{2008}).

\bibitem{Elste_Phys.Rev.B73/2006}
\bibinfo{author}{Elste, F.} \& \bibinfo{author}{Timm, C.}
\newblock \bibinfo{title}{{Transport through anisotropic magnetic molecules
  with partially ferromagnetic leads: Spin-charge conversion and negative
  differential conductance}}.
\newblock \emph{\bibinfo{journal}{Phys. Rev. B}} \textbf{\bibinfo{volume}{73}},
  \bibinfo{pages}{235305} (\bibinfo{year}{2006}).

\bibitem{Timm_Phys.Rev.B73/2006}
\bibinfo{author}{Timm, C.} \& \bibinfo{author}{Elste, F.}
\newblock \bibinfo{title}{{Spin amplification, reading, and writing in
  transport through anisotropic magnetic molecules}}.
\newblock \emph{\bibinfo{journal}{Phys. Rev. B}} \textbf{\bibinfo{volume}{73}},
  \bibinfo{pages}{235304} (\bibinfo{year}{2006}).

\bibitem{Misiorny_Phys.Rev.Lett.106/2011}
\bibinfo{author}{Misiorny, M.}, \bibinfo{author}{Weymann, I.} \&
  \bibinfo{author}{Barna\ifmmode~\acute{s}\else \'{s}\fi{}, J.}
\newblock \bibinfo{title}{{Interplay of the Kondo effect and spin-polarized
  transport in magnetic molecules, adatoms, and quantum dots}}.
\newblock \emph{\bibinfo{journal}{Phys. Rev. Lett.}}
  \textbf{\bibinfo{volume}{106}}, \bibinfo{pages}{126602}
  (\bibinfo{year}{2011}).

\bibitem{Onsager_Phys.Rev.38/1931}
\bibinfo{author}{Onsager, L.}
\newblock \bibinfo{title}{{Reciprocal relations in irreversible processes.
  II.}}
\newblock \emph{\bibinfo{journal}{Phys. Rev.}} \textbf{\bibinfo{volume}{38}},
  \bibinfo{pages}{2265} (\bibinfo{year}{1931}).

\bibitem{Misiorny2012Jul}
\bibinfo{author}{Misiorny, M.}, \bibinfo{author}{Weymann, I.} \&
  \bibinfo{author}{Barna{\ifmmode\acute{s}\else\'{s}\fi}, J.}
\newblock \bibinfo{title}{{Temperature dependence of electronic transport
  through molecular magnets in the Kondo regime}}.
\newblock \emph{\bibinfo{journal}{Phys. Rev. B}} \textbf{\bibinfo{volume}{86}},
  \bibinfo{pages}{035417} (\bibinfo{year}{2012}).

\bibitem{Wojcik_PhysRevB.91.134422/2014}
\bibinfo{author}{W\'ojcik, K.~P.} \& \bibinfo{author}{Weymann, I.}
\newblock \bibinfo{title}{Two-stage kondo effect in t-shaped double quantum
  dots with ferromagnetic leads}.
\newblock \emph{\bibinfo{journal}{Phys. Rev. B}} \textbf{\bibinfo{volume}{91}},
  \bibinfo{pages}{134422} (\bibinfo{year}{2015}).

\bibitem{Sahoo2005Nov}
\bibinfo{author}{Sahoo, S.} \emph{et~al.}
\newblock \bibinfo{title}{{Electric field control of spin transport}}.
\newblock \emph{\bibinfo{journal}{Nat. Phys.}} \textbf{\bibinfo{volume}{1}},
  \bibinfo{pages}{99--102} (\bibinfo{year}{2005}).

\bibitem{Merchant_Phys.Rev.Lett.100/2008}
\bibinfo{author}{Merchant, C.} \& \bibinfo{author}{Markovi\'{c}, N.}
\newblock \bibinfo{title}{{Electrically tunable spin polarization in a carbon
  nanotube spin diode}}.
\newblock \emph{\bibinfo{journal}{Phys. Rev. Lett.}}
  \textbf{\bibinfo{volume}{100}}, \bibinfo{pages}{156601}
  (\bibinfo{year}{2008}).

\bibitem{Gaass_Phys.Rev.Lett.107/2011}
\bibinfo{author}{Gaass, M.} \emph{et~al.}
\newblock \bibinfo{title}{{Universality of the Kondo effect in quantum dots
  with ferromagnetic leads}}.
\newblock \emph{\bibinfo{journal}{Phys. Rev. Lett.}}
  \textbf{\bibinfo{volume}{107}}, \bibinfo{pages}{176808}
  (\bibinfo{year}{2011}).

\bibitem{Martinek_Phys.Rev.Lett.91/2003_127203}
\bibinfo{author}{Martinek, J.} \emph{et~al.}
\newblock \bibinfo{title}{{Kondo effect in quantum dots coupled to
  ferromagnetic leads}}.
\newblock \emph{\bibinfo{journal}{Phys. Rev. Lett.}}
  \textbf{\bibinfo{volume}{91}}, \bibinfo{pages}{127203}
  (\bibinfo{year}{2003}).

\bibitem{Misiorny_Phys.Rev.B84/2011}
\bibinfo{author}{Misiorny, M.}, \bibinfo{author}{Weymann, I.} \&
  \bibinfo{author}{Barna\ifmmode~\acute{s}\else \'{s}\fi{}, J.}
\newblock \bibinfo{title}{{Influence of magnetic anisotropy on the Kondo effect
  and spin-polarized transport through magnetic molecules, adatoms, and quantum
  dots}}.
\newblock \emph{\bibinfo{journal}{Phys. Rev. B}} \textbf{\bibinfo{volume}{84}},
  \bibinfo{pages}{035445} (\bibinfo{year}{2011}).

\bibitem{Pasupathy_Science306/2004}
\bibinfo{author}{Pasupathy, A.} \emph{et~al.}
\newblock \bibinfo{title}{{The Kondo effect in the presence of
  ferromagnetism}}.
\newblock \emph{\bibinfo{journal}{Science}} \textbf{\bibinfo{volume}{306}},
  \bibinfo{pages}{86--89} (\bibinfo{year}{2004}).

\bibitem{Hauptmann_NaturePhys.4/2008}
\bibinfo{author}{Hauptmann, J.}, \bibinfo{author}{Paaske, J.} \&
  \bibinfo{author}{Lindelof, P.}
\newblock \bibinfo{title}{{Electric-field-controlled spin reversal in a quantum
  dot with ferromagnetic contacts}}.
\newblock \emph{\bibinfo{journal}{Nature Phys.}} \textbf{\bibinfo{volume}{4}},
  \bibinfo{pages}{373--376} (\bibinfo{year}{2008}).

\bibitem{bordoloi_double_2020}
\bibinfo{author}{Bordoloi, A.}, \bibinfo{author}{Zannier, V.},
  \bibinfo{author}{Sorba, L.}, \bibinfo{author}{Schönenberger, C.} \&
  \bibinfo{author}{Baumgartner, A.}
\newblock \bibinfo{title}{A double quantum dot spin valve}.
\newblock \emph{\bibinfo{journal}{Communications Physics}}
  \textbf{\bibinfo{volume}{3}}, \bibinfo{pages}{135} (\bibinfo{year}{2020}).
\newblock \urlprefix\url{http://www.nature.com/articles/s42005-020-00405-2}.

\bibitem{Legeza2008Sep}
\bibinfo{author}{Legeza, O.}, \bibinfo{author}{Moca, C.~P.},
  \bibinfo{author}{Toth, A.~I.}, \bibinfo{author}{Weymann, I.} \&
  \bibinfo{author}{Zarand, G.}
\newblock \bibinfo{title}{{We used the open-access Budapest Flexible DM-NRG
  code, http://www.phy.bme.hu/\~{}dmnrg/}} .

\bibitem{Toth2008Dec}
\bibinfo{author}{T{\ifmmode\acute{o}\else\'{o}\fi}th, A.~I.},
  \bibinfo{author}{Moca, C.~P.}, \bibinfo{author}{Legeza,
  {\ifmmode\ddot{O}\else\"{O}\fi}.} \&
  \bibinfo{author}{Zar{\ifmmode\acute{a}\else\'{a}\fi}nd, G.}
\newblock \bibinfo{title}{{Density matrix numerical renormalization group for
  non-Abelian symmetries}}.
\newblock \emph{\bibinfo{journal}{Phys. Rev. B}} \textbf{\bibinfo{volume}{78}},
  \bibinfo{pages}{245109} (\bibinfo{year}{2008}).

\bibitem{Anders2005}
\bibinfo{author}{Anders, F.} \& \bibinfo{author}{Schiller, A.}
\newblock \bibinfo{title}{Real-time dynamics in quantum-impurity systems: A
  time-dependent {N}umerical {R}enormalization-{G}roup approach}.
\newblock \emph{\bibinfo{journal}{Phys. Rev. Lett.}}
  \textbf{\bibinfo{volume}{95}}, \bibinfo{pages}{196801}
  (\bibinfo{year}{2005}).
\newblock
  \urlprefix\url{https://link.aps.org/doi/10.1103/PhysRevLett.95.196801}.

\bibitem{Weichselbaum_Phys.Rev.Lett.99/2007}
\bibinfo{author}{Weichselbaum, A.} \& \bibinfo{author}{von Delft, J.}
\newblock \bibinfo{title}{Sum-rule conserving spectral functions from the
  numerical renormalization group}.
\newblock \emph{\bibinfo{journal}{Phys. Rev. Lett.}}
  \textbf{\bibinfo{volume}{99}}, \bibinfo{pages}{76402} (\bibinfo{year}{2007}).

\end{thebibliography}

\end{document}